\documentclass[aps,prl,10pt,twocolumn,floatfix,superscriptaddress,showpacs,numerical,footinbib,longbibliography]{revtex4-1}
\usepackage[normalem]{ulem}
\usepackage{graphicx}
\usepackage{amsmath}
\usepackage{amssymb}
\usepackage{mathdots}
\usepackage{bbold}
\usepackage[usenames,dvipsnames,pdftex]{color}
\usepackage[%
  colorlinks=true,
  urlcolor=blue,
  linkcolor=blue,
  citecolor=blue
]{hyperref}

\newcommand{\ket}[1]{\left\vert #1 \right\rangle}

\newcommand{\pdag}{{\phantom{\dagger}}}

\newcommand{\mrm}[1]{\mathrm{#1}}

\renewcommand{\dag}{^{\dagger}}

\def\be{\begin{equation}}
\def\ee{\end{equation}}
\def\bea{\begin{eqnarray}}
\def\eea{\end{eqnarray}}

\def\ket#1{|#1\rangle }

\begin{document}

\title{{Charge excitation dynamics in bosonic fractional Chern insulators}}

\author{Xiao-Yu Dong }
\affiliation{Max-Planck-Institut f{\"u}r Physik komplexer Systeme, N{\"o}thnitzer Stra{\ss}e 38, 01187 Dresden, Germany}
\author{Adolfo G. Grushin}
\affiliation{Department of Physics, University of California, Berkeley, CA 94720, USA}
\affiliation{Institut N{\'e}el, CNRS and Universit{\'e} Grenoble Alpes, Grenoble, France}
\author{Johannes Motruk}
\affiliation{Department of Physics, University of California, Berkeley, CA 94720, USA}
\affiliation{Materials Science Division, Lawrence Berkeley National Laboratory, Berkeley, California 94720, USA}
\author{Frank Pollmann}
\affiliation{Max-Planck-Institut f{\"u}r Physik komplexer Systeme, N{\"o}thnitzer Stra{\ss}e 38, 01187 Dresden, Germany}
\affiliation{Technische Universit{\"a}t M{\"u}nchen, Physics Department T42, 85747 Garching, Germany}

\begin{abstract}
The experimental realization of the Harper-Hofstadter model in ultra-cold atomic gases has placed fractional states of matter in these systems within reach---a fractional Chern insulator state (FCI) is expected to emerge for sufficiently strong interactions when half-filling the lowest band.
The experimental setups naturally allow to probe the dynamics of this topological state, yet little is known about its out-of-equilibrium properties.
We explore, using  density matrix renormalization group (DMRG) simulations, the response of the FCI state to spatially localized perturbations.
After confirming the static properties of the phase we show that the characteristic, gapless features are clearly visible in the edge dynamics.
We find that a local edge perturbation in this model propagates chirally independent of the perturbation strength.
This contrasts the behavior of single particle models with counter-propagating edge states, such as the non-interacting Harper-Hofstadter model, where the chirality is manifest only for weak perturbations.
Additionally, our simulations show that there is inevitable density leakage from the first row of sites into the bulk, preventing a naive chiral Luttinger theory interpretation of the dynamics.
\end{abstract}

\maketitle

\textbf{Introduction.} Understanding the dynamical properties of strongly correlated quantum phases in dimensions higher than one still remains a difficult challenge in the vast majority of cases~\cite{Polkovnikov2011,Eisert2015}.
The lack of a complete paradigm originates from the inherent complexity of simulating the dynamics of strongly interacting quantum systems.
However, modern experiments~\cite{Bloch2005,Trotzky2012, Schreiber2015, Choi2016, tai2017microscopy} are now able to access time-dependent properties and thus the need to precisely characterize dynamical signatures of correlated phases is becoming pressing.
Among the most intriguing are scenarios in which topology joins in as an additional ingredient of the system.

A recent prominent example is the realization of the Chern insulator phase using ultracold atoms, both in a bosonic Harper-Hofstadter model~\cite{Harper1955,Hofstadter1976,Aidelsburger2013,Miyake2013} and the fermionic Haldane honeycomb model~\cite{Haldane1988,Jotzu2014}.
In both cases, periodically driving a lattice loaded with ultra cold atoms has been proven to show topological features~\cite{Aidelsburger2014,Jotzu2014},
as predicted by general theoretical arguments based on Floquet theory~\cite{Jaksch2003,Rahav2003,Dalibard2011,Goldman2016,Eckardt2017}.  
On-site interactions in the Harper-Hofstadter realization can drive the system into a bosonic Floquet fractional Chern insulator (FCI) state~\cite{Grushin2014,PhysRevB.91.245135,PhysRevA.93.043618,BERGHOLTZ2013,Neupert2015,Neupert2011,Tang2010,Sun2011,Regnault2011,Wu2012}, the bosonic periodically driven analog of the fractional quantum Hall (FQH) effect~\cite{Hafezi2007,Moeller2009,He2017,Gerster2017,Motruk2017,Hugel2017}.
Several protocols have been proposed to prepare this state and the phase diagram of the Harper-Hofstadter model for hardcore bosons has been established using various numerical methods~\cite{Sorensen2005,He2017,Motruk2017}.
Although this body of knowledge combined with proposals to detect chiral edge states~\cite{Spielman2013, Goldman2013, Goldman2012} hinted how to identify the existence of the FCI state in cold atomic experiments, simulations of dynamical signatures of this phase are still lacking.
However, the observation of time-dependent quantities in this system is possible, due to the high tunability of parameters and slow dynamics compared to the solid state, and necessary, due to the difficulty of probing transport quantities characterizing these states, such as the Hall conductivity.

\begin{figure}[ht]
 \includegraphics[width=\columnwidth]{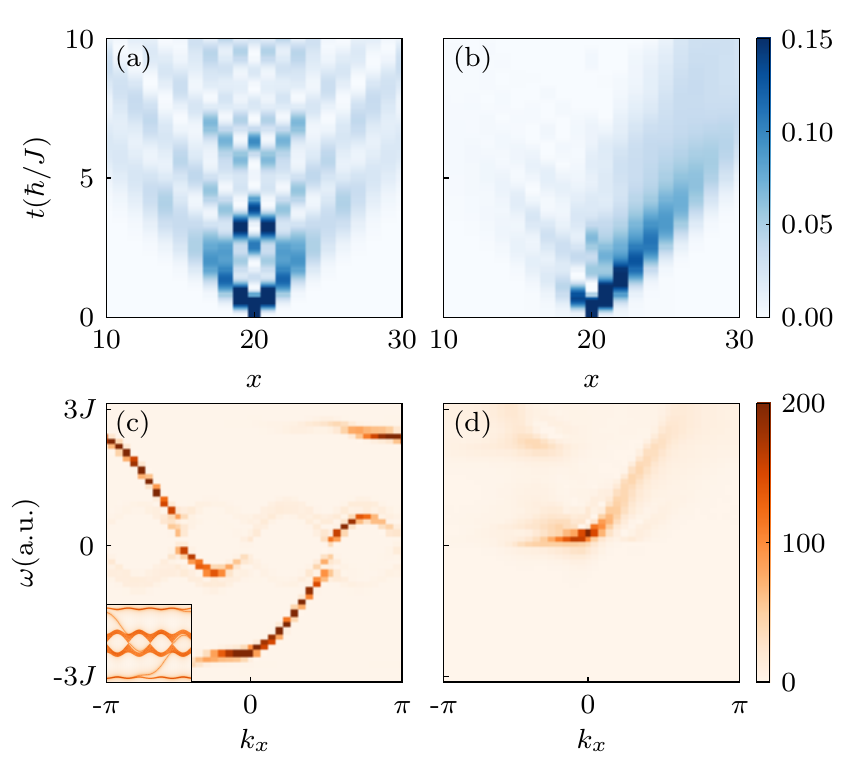}
 \caption{Time evolution of the {particle} density in the Harper-Hofstadter model after a particle has been created at the edge of an empty vacuum state with total filling $\nu=0$ (a) and in the interacting FCI state at $\nu=1/2$ of the lowest band (b). 
The spectral function $A(k_x,\omega)$ of the non-interacting case (c) reveals a prominent overlap with gapless edges states of both chiralities. The inset shows for comparison the density of states of the single-particle model. 
In contrast, the spectral function for the FCI (d) shows a single chiral mode. 
 \label{fig:mainfig}}
\end{figure}

In this Letter we address dynamical properties of the edge of the FCI phase of hardcore bosons at filling factor $\nu=1/2$ after local quenches using matrix-product state (MPS) based simulations.
We use the density matrix renormalization group (DMRG) method together with a recently introduced method \cite{Zaletel2015} that allows for the efficient simulation of the dynamical response function in two-dimensional systems \cite{Gohlke2017}. 
Our goal is to provide distinct dynamical signatures of the FCI phase which could be probed with current state of the art experiments.
By adding a particle at the edge we find a clear chiral propagation of the FCI gapless edge modes, characteristic for such phases (see Fig.~\ref{fig:mainfig}).
Moreover, this protocol provides a simple distinction between an emergent Laughlin state and a non-interacting Chern insulator (CI) that hosts multiple edge modes of opposite chirality.
The latter shows no chiral asymmetry while the chirality in the former case is clearly visible.
The reason is that a generic perturbation in the non-interacting case mixes edge states with opposite chirality  while the chirality in the Laughlin case is protected by the many-body bulk state.
As an experimentally relevant example we study the $\phi=\pi/2$ Harper-Hofstadter model at total filling $1/8$~\cite{Harper1955,Hofstadter1976,Aidelsburger2013,Miyake2013}  and propose a protocol that applies a local trap at the edge to distinguish the FCI state by varying the trap strength.
%
%---------figure------------------
%----------------------------------

\textbf{Static properties.}  We consider the Harper-Hofstadter Hamiltonian \cite{Harper1955,Hofstadter1976}%
\begin{equation}
\begin{split}
\label{eq:hamiltonian}
H&= -J \sum_{\langle ij \rangle} \left( e^{i \phi_{ij}} a_i\dag a_j\pdag + {\mrm h.\mrm c.} \right)
\end{split}
\end{equation}
on a square lattice with a magnetic flux of $\phi=\pi/2$ per plaquette. Here  $a_i\dag$ ($a_i$) creates (annihilates) a hardcore boson on site $i$.
The single-particle spectrum of $H$ has four bands (see Fig.~\ref{fig:mainfig}c, inset) with the central bands touching at four Dirac points. The model is characterized, from top to bottom, by three Chern numbers $C_i = \pm (1,-2,1)$ where the sign is determined by the sign of $\phi$.
%
%---------figure------------------
\begin{figure}
 \includegraphics[width=\columnwidth]{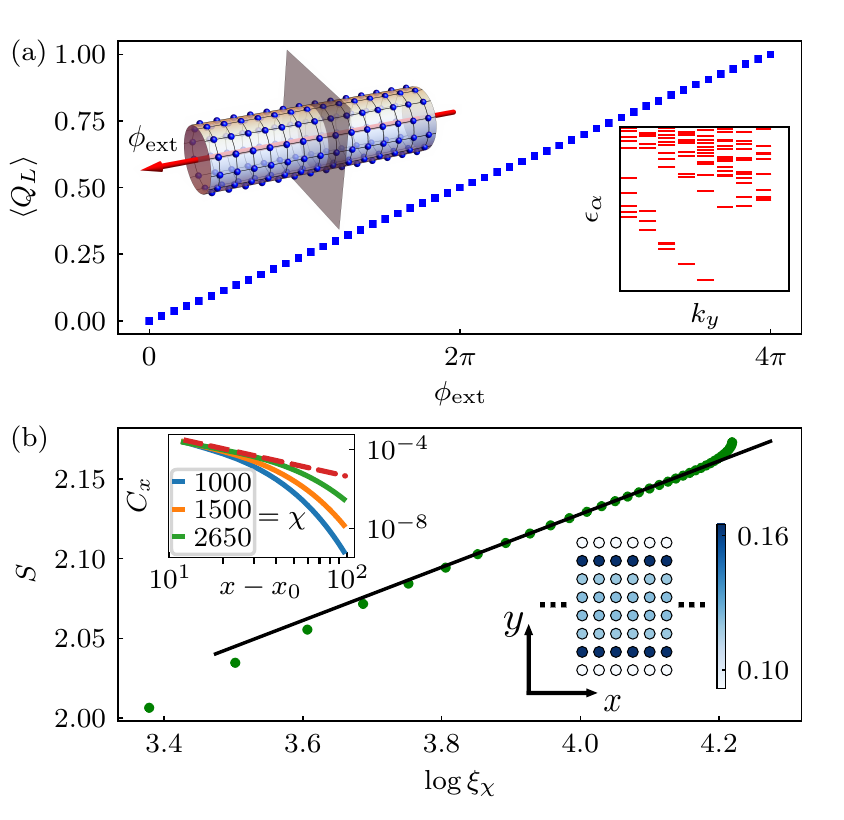}
 \caption{Static properties of the bosonic $\nu=1/2$ fractional Chern insulator. Panel (a) shows the pumping of a charge per two flux periods as expected for a $\nu=1/2$ FCI state. The inset shows the entanglement spectrum of the zero charge-sector. The low lying states satisfy the expected {conformal field theory (CFT)} counting $\{1,1,2,3,5,\dots\}$ (see also \cite{SM}). {All data in (a) are calculated in an infinitely long cylinder with $L_y=8$.}
 Panel (b) shows the scaling of the entanglement entropy $S$ as a function of the correlation length $\xi_\chi$ for an infinite strip. The slope of $c/6$ determines the central charge of the edge theory $c=1$. The lower right inset displays the real space charge density of a strip configuration, which is infinite in the $x$-direction and finite in the $y$-direction with $L_y=8$. {The upper left inset shows the ground state correlation function $C_x = \langle a_x a^{\dag}_{x_0} \rangle$ on the edge versus $x-x_0$ of an infinite strip with $L_y=10$. 
The dashed line $\propto (x-x_0)^{-2}$ follows the Luttinger liquid theory prediction.
}
}
 \label{fig:static}
\end{figure}
%----------------------------------
We start {by verifying that $H$ indeed hosts a} $\nu=1/2$ Laughlin state in agreement with previous results~\cite{Hafezi2007,Moeller2009,CV13,He2017,Gerster2017,Motruk2017}.
For this we simulate Hamiltonian~\eqref{eq:hamiltonian} on an infinite cylinder of circumference $L_{y}$ and total filling $1/8$ with DMRG, which enforces the half-filling of the lowest Chern band.
The results are summarized in Fig.~\ref{fig:static} and \cite{SM} which confirm the topological nature of the state. 
We find a quantized Hall conductivity of $\nu=1/2$, the characteristic structure in the entanglement spectrum, {the static correlation function on the edge that approaches the prediction of Luttinger liquid theory with increasing bond dimension ($\chi$),} and a central charge of $c=1$ for the edge theory through a finite entanglement scaling~\cite{Tagliacozzo2008,Pollmann2009} when considering an infinite strip geometry.
The latter quantity shows that the DMRG simulations on the infinite strip reproduce the expected critical behavior at the edge {and that edge overlap is negligible for our choices of $L_y \geq 8$} \cite{calabrese2004entanglement,Li2008}.
%
%
%%%%%%
%
%
%Protocols-----------------
%
%
%%%%%

\textbf{Evolution of an added particle at the edge.}
Having established the presence of the many-body FCI state at 1/8 filling,
we will now focus on the dynamical response.
The results are shown in Fig.~\ref{fig:mainfig} and reveal characteristic differences between the single-particle and FCI case.
We first investigate the single-particle case and consider the system on a strip geometry with open (periodic) boundary conditions along $y$ ($x$).
A particle is created at the edge of an empty lattice by acting on it with an $a\dag$ operator and the resulting state is then evolved in time.
Figure~\ref{fig:mainfig}a shows the time evolution of the particle density on the edge which exhibits no chirality; this can be understood by the following reasoning. 
The single-particle spectrum of the model (shown in the inset of Fig.~\ref{fig:mainfig}c) possesses two dispersing mid-gap modes at different energies of opposite chiralities.
These connect the central band ($C=-2$) to the top and bottom bands (C = 1) and \textit{both} modes are exponentially localized at the edge of the finite strip. 
When creating a single particle at one edge, the state has overlap with both edge modes since both have support on the edge where the particle is created leading to the symmetric dispersion of the particle density.
To verify the above interpretation, we compute the spectral function $A(k_x,\omega)$ as the Fourier transform  in space and time of the dynamical correlation function {$C_x(t)=\langle a_x(t) a^{\dag}_{x_0}(t_0) \rangle$} for momentum $k_x$ along the edge at frequency $\omega$ shown in Fig.~\ref{fig:mainfig}c.
When compared to the energy spectrum of $H$ (Fig.~\ref{fig:mainfig}c, inset), the spectral function highlights the fact that both mid-gap chiral states have overlap with the created particle, explaining the achiral behavior observed when simulating the time evolution.

For the interacting case, we consider an infinite strip geometry at $\nu = 1/2$ filling of the lowest band to prepare the system in an FCI ground state $\ket{\Psi^{\mrm{FCI}}_{\mrm{strip}}}$ (see {lower right} inset of Fig.~\ref{fig:static}b).
We again create a particle at the edge to obtain the state $\ket{\Psi_{i}} = a^{\dagger}_i \ket{\Psi^{\mrm{FCI}}_{\mrm{strip}}}$.
We then simulate the time evolution of $\ket{\Psi_{i}}$ under the Hamiltonian $H$ using a matrix-product operator based time evolution method \cite{KPM11,Zaletel2015,Gohlke2017, SM}.
Unlike in the free particle case, the propagation of the density is chiral (Fig.~\ref{fig:mainfig}b) consistent with the single chiral branch in the spectral function (Fig.~\ref{fig:mainfig}d).
In the FCI state, the emergent chirality is protected by the topology of the many-body wave function in the bulk and thus it is more robust than the single-particle case.

%---------figure------------------
\begin{figure}
 \includegraphics[width=\columnwidth]{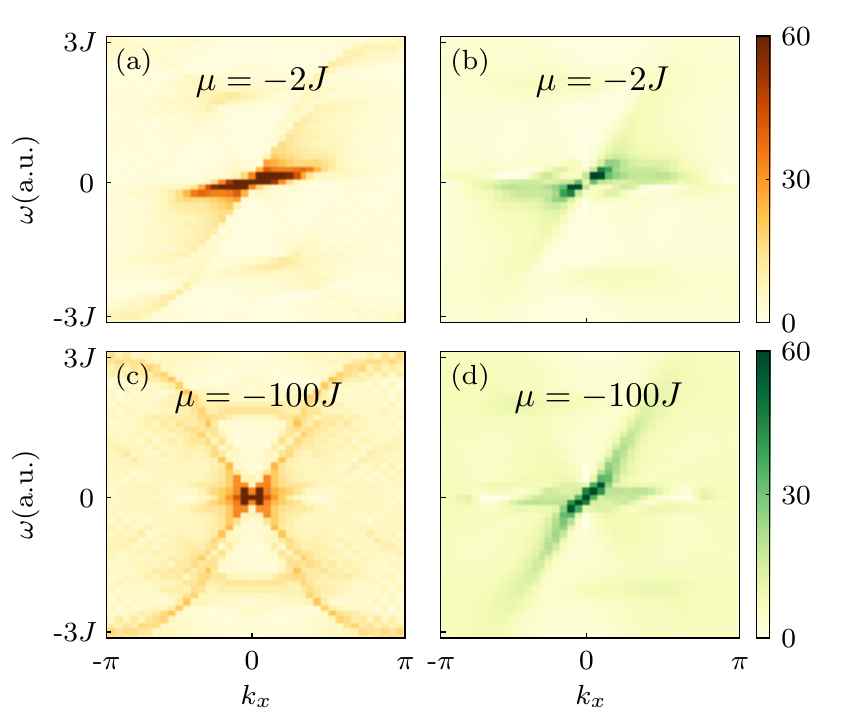}
 \caption{Fourier transformation of the {particle} density evolution for a shallow ($\mu=-2J$) and a deep trap ($\mu=-100J$) localized at an edge site of the single-particle (a,c) and FCI case (b,d).}
 \label{fig:traps}
\end{figure}
%----------------------------------

\textbf{Trapping potential and dynamics.} With the insight gained previously it is possible to devise a protocol closer to what is experimentally realizable.
In cold atomic systems, lasers are used to control the local density of particles, making it possible to create a local trapping potential of varying strength of the form \cite{Bakr2009,Hilker484}
\be
\label{eq:pert}
H_{\mu} = \mu a^{\dagger}_{i}a_{i}.
\ee
We again restrict $i\in \mrm{edge}$ and compare the response of the single-particle case and the hardcore boson $\nu=1/2$ FCI state as a function of $\mu$.
Our results are shown in Fig.~\ref{fig:traps} where we plot the Fourier transform of the {particle} density evolution on the edge, {i.e. $\int dx \int dt a^{\dag}_x(t)a_x(t)$ with $x$ on the edge.}  (see also \cite{SM}).
For the single-particle scenario in Figs.~\ref{fig:traps}a and c, we fill the lowest energy state of Hamiltonian~\eqref{eq:hamiltonian} with one particle in the presence of a finite $\mu$ and then time evolve the resulting state with a quenched Hamiltonian by abruptly switching off the local potential.
As a function of the trapping potential, the Fourier transform shows a non-symmetric (symmetric) structure corresponding to a chiral (achiral) density evolution for a shallow (deep) trap (see Fig.~\ref{fig:traps}a and c).
This difference originates from the fact that for a given $\mu$, the evolving state can only explore
a sub-set of the band structure.
If $\mu$ is smaller than the gap between lowest and central band, then time evolution allows to explore states only within one chiral edge state and thus exhibits chiral behavior.
If $\mu$ is large compared to the total band width, the initial state has overlap with the entire spectrum after switching off the potential. 
As discussed previously for Fig.~\ref{fig:mainfig}c, these states include two chiral modes of opposite chirality, and thus the chiral propagation disappears (Fig.~\ref{fig:traps}c).

For the interacting case, we find the ground state for finite $\mu$ on an infinite strip at total filling $1/8$ with an extra particle using DMRG and subsequently let the state evolve under the quenched ($\mu=0$) Hamiltonian. 
The $\nu=1/2$ FCI state is a topologically ordered many-body state, and thus the single-particle band structure arguments do not apply.
The evolution stays chiral for arbitrary trapping potential strength, as we observe in Fig.~\ref{fig:traps}b and d.
In this case, the many-body state dictates the excitations at the edge which prove to be chiral in one direction.
Taken together, the chiral evolution and the insensitivity to the trapping potential can be probed as an experimental signature of the $\nu=1/2$ FCI state in this model, and is therefore one of the main results of this Letter.

\begin{figure}
 \includegraphics[width=\columnwidth]{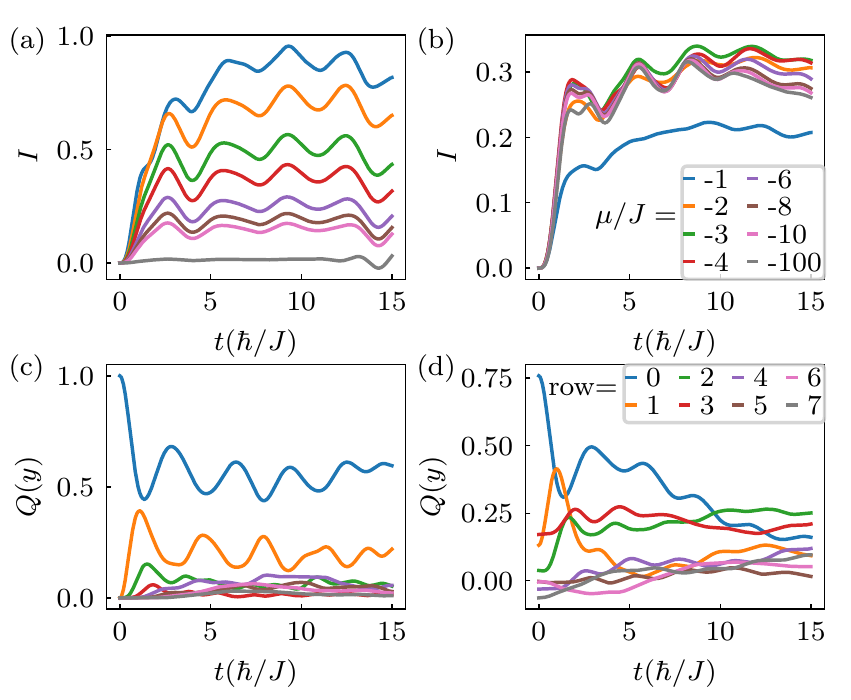}
 \caption{The time evolution of the imbalance $\mathcal{I}=N_R-N_L$ between the total particle density on the right and left part of site $i$ for the single-particle (a) and FCI (b) case for different values of the perturbation $\mu$ in units of the hopping $J$. For single particle (FCI) case the imbalance decreases (saturates) with increasing perturbation strength.  Panels (c) and (d) show the total density per row as a function of time for the single-particle and the interacting scenario, respectively for $\mu=-100J$, showing a sizable leakage of particle density into the bulk.  The legend indicates the row number in the $y$-direction in the geometry of the inset of Fig.\ref{fig:static}b.}
 \label{fig:leak}
\end{figure}

In order to quantify the dependence of the chirality on the value of $\mu$, we define the imbalance $\mathcal{I}=N_{R}-N_{L}$ of the total particle number on the edge to the left and right of site $i$ during the time evolution in Figs.~\ref{fig:leak}a and b.
The single-particle case is shown in Fig.~\ref{fig:leak}a and the imbalance decreases with increasing the absolute value of $\mu$ consistent with the explanation above.
For a very deep trap, the difference is almost zero, which denotes achiral behavior as expected.
In contrast, the chiral behavior of the interacting topologically ordered state persists even for a very deep trap as shown in Fig.~\ref{fig:leak}b.

%%%%%%
%
%%%%%
%Leak-----------------
%
%
%%%%%

%---------figure------------------
%----------------------------------

\textbf{Towards a chiral Luttinger liquid description.} It is tempting to connect our previous analysis with the chiral Luttinger liquid of quantum Hall edge states~\cite{Wen1990}.
For a Laughlin state at a filling $\nu=1/m$, this description predicts that the spectral function and the density of states behave as \cite{Wen1990,SM}
\begin{eqnarray}
\label{eq:chiral-lutt}
A(k,\omega) \propto (\omega+v k)^{m-1}\delta(\omega-vk),\; \;
N(\omega)\propto \omega^{m-1},
\end{eqnarray}
where 
$v$ is the velocity of the edge state.
A direct measurement of $A(k,\omega)$ or $N(\omega)=\int_k dk A(k,\omega)$ could be used to extract $m$, which would be solid evidence for the presence of the FCI state in experiment.
Such analytical spectral function could be in principle compared directly to our numerical spectral function in Fig.~\ref{fig:mainfig}d.
However, this exercise reveals two potential problems that experiments may face to extract $m$.
First, the main differences between a trivial edge state and a chiral Luttinger liquid will be most drastic at longer times, or smaller $\omega$.
This region is however the most elusive numerically, due to entanglement growth, and experimentally, due to heating and particle loss.
Second, particles created at the edge have a finite overlap with bulk states as the correlation length is finite. 
Consequently, the particle will diffuse into the bulk at longer times, making it difficult to resolve the low-energy (long time) behavior of the edge. 
We have numerically observed that a sizable part of the edge density is lost into the bulk.
Our results are shown in Fig.~\ref{fig:leak}c and d where we plot the average density per row for the free and interacting cases respectively when a particle is added at the edge (row $y=0$).
In both cases, we find that there is a leakage of density to the bulk, and the physical edge (i.e., the first row of sites) does not behave as an isolated liquid.
In the single-particle case of Fig.~\ref{fig:leak}c, the particle density stabilizes after an initial drop, features that may be explained by the high overlap of the initial state with the exponentially localized edge eigenstates of the spectrum.
The interacting case in Fig.~\ref{fig:leak}d suffers from a more severe particle loss to the bulk of the system. 
We have attempted several protocols to decrease such a leakage.
First, by increasing the width of the strip  $L_{y}=4,8,12$ {we find no} appreciable change in the density loss.
This is consistent with the fact that for any finite width, the interactions between the two edge states are marginal for $\nu=1/2$~\cite{Wen1990}.
Second, we have tried to confine the chiral edge modes with an additional negative chemical potential localized at the edge.
We observed that although it reduces the leakage at long times, sufficient density is lost at short times to prevent a comparison with Eq.~\eqref{eq:chiral-lutt}.
Third, we find that the leakage is reduced by choosing $J_{y}/J_{x}<1$.
By studying the static properties as a function of $J_y$, we have checked that  the FCI phase with $\nu=1/2$ is stable up to strong anisotropies \cite{He2017,Hugel2017}. 
The smaller leakage as $J_y$ decreases indicates that the correlation between the edge state and the bulk states in {the $y$-direction} is the main source for particle loss. 
%

%%%%%%
%
%
%Discussion-----------------
%
%
%%%%%
\textbf{Conclusions.} In this paper, we have studied the dynamical properties
of a bosonic fractional Chern insulator edge under local perturbations using the infinite density matrix renormalization group.
We have dynamically established the chirality of the $\nu=1/2$ bosonic FCI state emergent in the Harper-Hofstadter model at 1/8 total filling, a relevant example 
for current cold atom experiments. 
We found that in the fractional Chern insulating phase a generic edge perturbation in this model propagates chirally, while the chirality in the single-particle case is only visible for weak perturbations, up to the order of the gap between the lowest and central band.
This distinction can be carried over to Chern insulating models which host chiral edge modes with opposite chirality coexisting at a given edge, a common instance for multiband models, such as the Harper-Hofstader model.
In contrast, two band Chern insulator models, such as the Haldane model~\cite{Haldane1988} realized experimentally~\cite{Jotzu2014} fall outside this category, and our approach is in principle not sufficient to distinguish  the CI from the FCI state.
However, in these models, if $\mu$ is larger than the single particle band gap, bulk excitations will be created, which will introduce larger noise to the chiral signal~\cite{Grushin2016}
than in their FCI counterparts, where the gap is set by the strong interaction energy scale.
Our simulations also showed that there is inevitable density leakage into the bulk, preventing a naive chiral Luttinger theory interpretation of the dynamics.

Recently, a related example of the interplay between interactions and topology causing chiral dynamics was experimentally observed on a ladder system underlining the relevance of our results to ongoing experiments~\cite{tai2017microscopy}.
This experiment employed a box-like confining potential that brings in line with our numerical simulations and which circumvents the effects of harmonic confinement~\cite{Buchhold2012, Goldman2013identifying}.
A different realistic alternative is the engineering of sharp interfaces~\cite{Goldman2016a}. 

Our work highlights that in realistic experimental set-ups a richer dynamical behavior beyond a naive 1D Luttinger liquid behaviour should be expected in fractional Chern insulators. 
It is triggered by an unforeseen density leakage from the first row of sites and the insensibility to the energy scales set by a perturbation localized to the edge, emphasizing the need for further studies of dynamics of fractional Chern insulators.

\textbf{Acknowledgements} 
We thank N. Goldman, F. Grusdt, M. Kolodrubetz, N. Regnault and R. Vasseur for fruitful discussions and suggestions. 
A. G. G was supported by the Marie Curie Programme under EC Grant agreement No.\ 653846. 
J. M. acknowledges funding by TIMES at Lawrence Berkeley National Laboratory supported by the U.S. Department of Energy, Office of Basic Energy Sciences, Division of Materials Sciences and Engineering, under Contract No.\ DE-AC02-76SF00515 and through DFG research fellowship MO 3278/1-1. 
F. P. acknowledges support from DFG through Research Unit FOR 1807 with Grant No. PO 1370/2-1, the Nanosystems Initiative Munich (NIM) by the German Excellence Initiative, and the European Research Council (ERC) under the European Union’s Horizon 2020 research and innovation program (grant agreement no. 771537).

\appendix

\section{The Hamiltonian of the Hofstadter model}

The Hamiltonian of the bosonic Harper-Hofstadter model on a 2D square lattice is
\begin{eqnarray}
\label{eq:H}
\label{eq:HH}
&&H_{\mrm{HH}} = -\bigg[\sum_x(J_x\sum_{y=1}^{L_y}a^{\dag}_{x+1,y}a_{x,y}e^{-iy\pi/2}\\
\nonumber
 &&\qquad+J_y\sum_{y=1}^{L_y-1} a^{\dag}_{x,y+1}a_{x,y} +P_{bc}J_ya^{\dag}_{x,1}a_{x,L_y})+\mathrm{H.c.}\bigg]
 \label{eq:int}
\end{eqnarray}
where $(x,y)$ are the coordinates of the lattice sites and $a^{\dag}_{x,y}$ ($a_{x,y}$) is the creation (annihilation) operator of a hardcore boson at site $(x,y)$. 
Physically, the hardcore constraint is achieved by an implicit onsite interaction term $ U/2 \sum_{x,y} n_{x,y} (n_{x,y}-1)$.
Taking the interaction $ U \rightarrow \infty $ the particle number operator at site $(x,y)$, given by $n_{x,y}=a^{\dag}_{x,y}a_{x,y}$,
is constrained to $n_{x,y}=0,1$.
The parameters $J_x$ and $J_y$ quantify the hopping in the $x$- and $y$-direction and the parameter $P_{bc}=0,1$ respectively sets open or periodic boundary conditions along the $y$-direction.
The system number of lattice sites in $y$-direction is given by the integer $L_y$, while in the $x$-direction, we allow the lattice to be finite or infinitely long. 
The model is written in the Landau gauge with $(A_x=-y\pi/2,A_y=0)$. 
Each plaquette is pierced by a flux of $\pi/2$, and the magnetic unit cell is $1\times 4$ with four square plaquettes along the $y$-direction.
Note that for the computation of the entanglement spectrum shown in the inset of Fig.~\ref{fig:static}a (main text), we change the gauge to $(A_x=0,A_y=-x\pi/2)$. This modifies the unit cell to $4 \times 1$ sites and therefore leads to $8$ momentum values around the cylinder.

\section{Details of the calculations of static properties}
In this section we present the details leading to the static properties of Hamiltonian Eq.~\eqref{eq:H} of hardcore bosons at 1/8 total filling of an infinitely long cylinder with circumference $L_y=8$. 
To this end we use a DMRG algorithm~\cite{McCulloch2008,Kjall2013} in a matrix product state (MPS) representation characterized by a bond dimension $\chi$ to find the ground-state of the system.
The ground state energy converges with increasing  $\chi$ to the energy $\sim -0.34117/J$ with $J = J_x = J_y = 1$. 
Both the correlation length along the cylinder and the entanglement energy converge to a constant with increasing $\chi$ which indicates that the bulk state of the system has a finite energy gap and may therefore be faithfully represented by an MPS with finite $\chi$. 
The correlation length in units of the lattice constant is $\sim 1$ along the cylinder indicating that there is no long range Landau order parameter. 

We look at different quantities in order to verify the topological properties of the state.
A hallmark of the $\nu=1/2$ Laughlin state is the quantized Hall conductivity $\sigma_{xy} = e^2/2h$ which we calculate in iDRMG by flux insertion~\cite{Laughlin1981,Zaletel2014,Grushin2015}.
To this end, we cut the cylinder into two semi-infinite halves and write the ground state wave function $\ket{\psi_0}$ in a Schmidt decomposition as
\begin{eqnarray}
 \ket{\psi_0} = \sum_{\alpha} \Lambda_{\alpha}|\alpha\rangle_{\mathrm{L}} \otimes |\alpha\rangle_{\mathrm{R}},
\end{eqnarray}
where $|\alpha\rangle_{\mathrm{L}}$ and $|\alpha\rangle_{\mathrm{R}}$ are states on the left and right half, respectively.
These Schmidt states can be assigned charge values $Q^{\alpha}_{L/R}$ so that the average charge $\langle Q_{L/R} \rangle$ of the left/right half of the system is given by
\begin{eqnarray}
 \langle Q_{L/R} \rangle = \sum_{\alpha} \Lambda^2_{\alpha} Q^{\alpha}_{L/R}.
\end{eqnarray}
When adiabatically inserting a flux $\phi_{\mathrm{ext}}$ through the system along the cylinder axis (see inset of Fig.~\ref{fig:static}a of the main text), a charge of $\sigma_{xy} \cdot (\phi_{\mathrm{ext}}/\phi_0)$ flows across the cut which we can calculate by monitoring the charge increase in the left half of the system.
Here, $\phi_0$ is the elementary flux quantum.
Fig.~\ref{fig:static}a in the main text shows that one unit of charge is pumped when inserting a flux of $4 \pi$, corresponding to a Hall conductivity of $\sigma_{xy} = 1/2$. In practice, the flux insertion is done by twisting the boundary conditions, i.e. setting $P_{bc} = e^{i \phi_\mathrm{ext}}$.

The second characteristic of the topologically ordered state consists of the structure of the entanglement spectrum.
Since the cylinder exhibits translation symmetry along the $y$-direction, the Schmidt states are eigenstates of the momentum $k_y$ in the $y$-direction and the corresponding entanglement energy levels $\epsilon_{\alpha} = - 2 \ln \Lambda_{\alpha}$ may be labelled with $k_y$.
In the inset of Fig.~\ref{fig:static}a of the main text, we plot the momentum resolved entanglement spectrum of the charge zero sector. 
For the low lying spectrum, we observe a structure governed by the conformal field theory (CFT) of the edge with a level counting of $\{1,1,2,3,5,\dots\}$ which indicates the presence of the virtual edge when cutting the system~\cite{Li2008, Grushin2015}.

As a final signature of the topological state, we compute the central charge of the edge CFT. 
To this end, we consider a strip geometry (shown in the inset of Fig.~\ref{fig:static}b of the main text) using open boundary conditions in the $y$-direction with $P_{bc}=0$. 
This method exploits the relation between the entanglement entropy $S$ 
\begin{eqnarray}
S = -\sum_{\alpha} \Lambda^2_{\alpha}\log\Lambda^2_{\alpha}
\end{eqnarray}
and the correlation length $\xi_\chi$. 
{While the physical correlation length at a critical point 
is infinite, a variationally optimized MPS with finite
bond dimension will always have a finite correlation
length $\xi_\chi$. 
For conformally invariant 
critical points it can be shown that $\xi_\chi \propto \chi^{\kappa}$ 
with $\kappa$ depending on the central charge~\cite{Tagliacozzo2008,Pollmann2009}. 
Moreover, the entanglement entropy between 
two halves of a large one-dimensional system close to a critical point 
is given by 
\begin{eqnarray}
S=\frac{c}{6}\log(\xi_\chi/a).
\label{Eq:3}
\end{eqnarray}
Here, $c$ is the \emph{central charge}
of the conformal field theory describing the critical point 
and $a$ is a short-distance length scale, in our case the lattice spacing which we set to be unity. 
Thus, a convenient method to extract 
the central charge is to perform iDMRG simulations with
different bond dimensions $\chi$ and then fitting the logarithmic
growth of the entanglement entropy $S$ as function of $\xi_\chi$.
From the topological properties of the bulk, we expect that the edge state is described by a chiral boson CFT 
with central charge $c=1$ which we compute by finite entanglement scaling shown in Fig.~\ref{fig:static}b of the main text~\cite{Tagliacozzo2008,Pollmann2009}. 
}

The black solid line in Fig.~\ref{fig:static}b in the main text has the slope $1/6$ confirming that $c=1$.
The deviations at large $\xi_\chi$ can be explained by noting that, in a system with a finite width, the two edges states may couple weakly to induce a small energy gap, and the true correlation length will not be infinite. 
When the bond dimension $\chi$ of the MPS is very large, the correlation length will stop growing, while the entanglement entropy still grows with increasing $\chi$. 
Therefore, at large $\xi_\chi$, we expect the relation in Eq. (8) in the main text not to hold; the coefficient between $S$ and $\log\xi_\chi$ will be larger than $c/6$, consistent with our numerical result (see Fig.~\ref{fig:static}b, main text).

{\section{Details of the calculation of the dynamical properties}

For the dynamical simulations we use a method introduced in \cite{KPM11} combined with the MPO based time evolution described in detail in~\cite{Zaletel2015,Gohlke2017}. 
% %
For the first protocol, we take a segment of several MPS unit cells (rings of sites around the cylinder) together with the boundary conditions from the unperturbed environment, act with $a_i^{\dag}$ on site $i$ in the center of the segment, and time-evolve this state $\ket{\Psi_{i}}$.
As long as the light cone of the perturbation is smaller than the size of the segment, this provides an efficient and reliable method to obtain dynamical response functions.
In the case of the second protocol with additional chemical potential, we compute the ground state of the Hamiltonian including the term $\mu_i n_i$ with the MPS unit cell having the size of the entire segment and then perform the time evolution under the Hamiltonian without local chemical potential term.
The maximum bond dimension we use is $\chi=200$, and the time step is $dt=0.02$.
We have checked that the quantities we present have converged with respect to bond dimension and time step size.}

\section{Chiral Luttinger liquid theory and comparison to DMRG}

In this section we review the main properties of a chiral Luttinger liquid ($\chi$LL)~\cite{Wen1990} that describes a 1D critical system that propagates only in one direction.
It is considered to be a good description of the edge of a fractional quantum Hall state.
For a Laughlin state of filling fraction $\nu=1/m$, where $m>0$ is an odd integer for fermionic system, and an even integer for bosonic systems, the edge state  is described by the Lagrangian
\begin{eqnarray}
\mathcal{L}=\frac{m}{4\pi}\partial_x\phi_{\mathrm{L}}(\partial_t\phi_{\mathrm{L}}-v\partial_x\phi_{\mathrm{L}})
\end{eqnarray}
where $\phi_{\mathrm{L}}$ is a free phonon field with a propagator $\langle \phi_{\mathrm{L}}(x,t)\phi_{\mathrm{L}}(0,0)\rangle=-\nu\ln(x-vt)$, and $v$ is velocity of the edge state. 
This Lagrangian can be obtained from a Chern-Simons theory in a finite system by imposing gauge invariance. 
The field operator of the elementary particle is $\Psi \propto e^{i\phi\sqrt{m}}$, and satisfies $\Psi(x)\Psi(x')=(-1)^{m}\Psi(x')\Psi(x)$.
Therefore, for $\nu=1/2$ we considered, $\Psi(x)$ is a bosonic operator. The Green's function is then given by
\begin{eqnarray}
G(x,t) &=& -i\langle \Phi_0|T(\hat{\Psi}(x,t)\hat{\Psi}^{\dag}(0,0))|\Phi_0\rangle,\nonumber\\
&\propto & \frac{1}{(x-vt)^m}
\end{eqnarray}
and the Fourier transform of the Green's function reads
\begin{eqnarray}
G(k,\omega) \propto \frac{(\omega+vk)^{m-1}}{\omega-vk+i0^{\dag}\mathrm{sgn}(\omega)}.
\end{eqnarray}
Thus, the spectral function has the form
\begin{eqnarray}
A(k,\omega)&=&-\frac{1}{\pi}\mathrm{Im}G(k,\omega) \nonumber\\
&\propto &\mathrm{sgn}(\omega)(\omega+v k)^{m-1}\delta(\omega-vk),\label{Eq:2}
\end{eqnarray}
and the density of states is given by
\begin{eqnarray}
N(\omega) = \int dk A(k,\omega) \propto |\omega|^{m-1}.
\end{eqnarray}

\section{Real space time evolution of the edge state density distribution}
\begin{figure}[t]
 \includegraphics[width=\columnwidth]{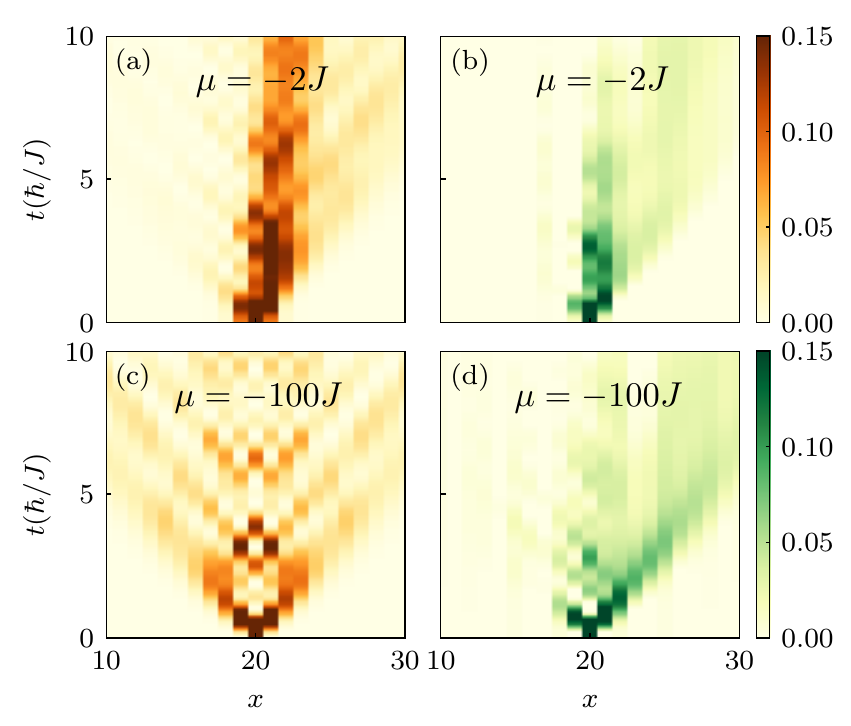}
 \caption{Edge state particle density evolution for a shallow ($\mu=-2J$) and a deep trap ($\mu=-100J$) localized at an edge site of the single-particle (a,c) and FCI case (b,d).}
 \label{fig: trap_real}
\end{figure}

\begin{figure}[t]
 \includegraphics[width=\columnwidth]{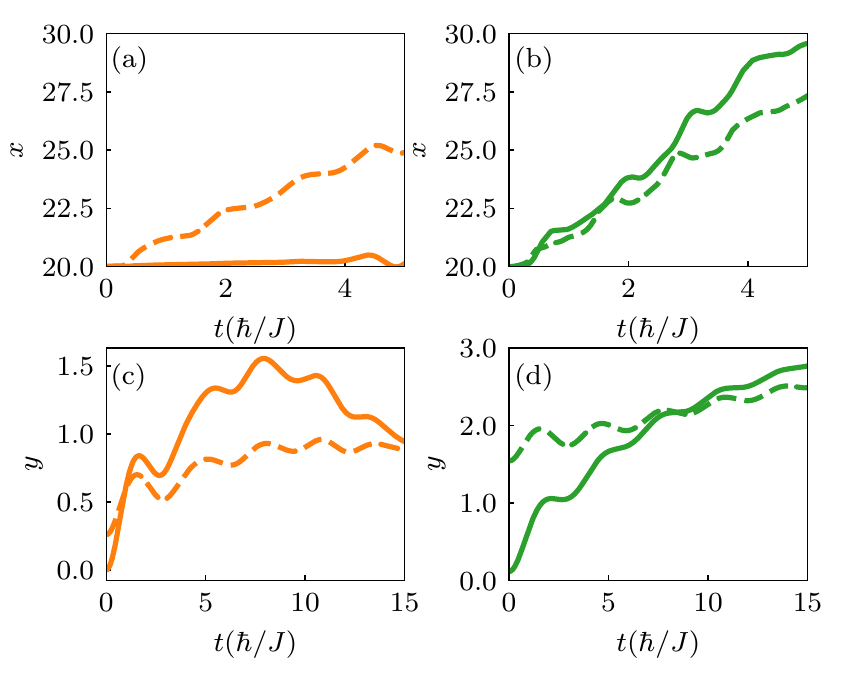}
 \caption{{Time evolution of the center-of-mass in $x$-direction on the edge line and $y$-direction for the single-particle (a,c) and FCI (b,d) phase. Solid (dashed) lines correspond to the cases with $\mu = -100J$ ($\mu = -2J$).}}
 \label{fig:COM}
\end{figure}

In Fig.~\ref{fig: trap_real} we give the time evolution of the {particle} density distribution {on the edge} in real space corresponding to the Fourier transformation in Fig.~\ref{fig:traps} of the main text.
In the single particle case, a shallow or deep trap leads to a chiral  or achiral propagation of the non-interacting edge state, respectively. 
The interacting edge state propagates chirally insensitive to the trap scale, consistent with the discussion presented in the main text.
{We also plot the time evolution of the center-of-mass in $x$-direction on the edge line and in $y$-direction with $\mu = -100J$ (solid lines) and $-2J$ (dashed lines) in Fig.~\ref{fig:COM}, which are measurable in current experiments \cite{Aidelsburger2014}.}
%---------figure------------------
%----------------------------------

%
%---------figure------------------
%----------------------------------

\begin{figure}
 \includegraphics[width=\columnwidth]{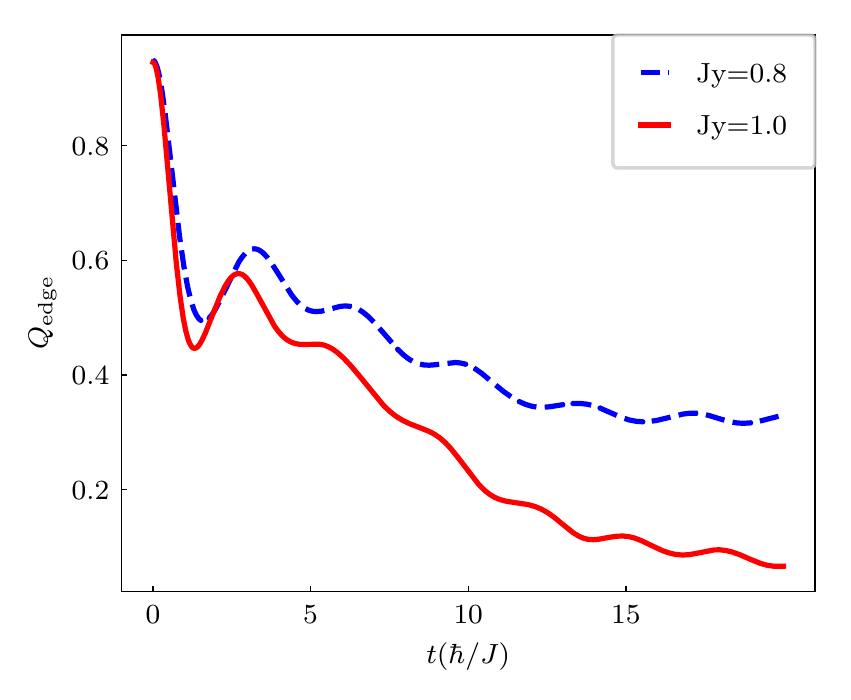}
 \caption{{Comparison of the total number of particles on the edge line $Q_{\mathrm{edge}}$ as a function of time $t$ in the interacting FCI phase for $J_y=0.8J$ and $1.0J$. }}
 \label{fig:diff_Jy}
\end{figure}

{\section{Comparison of particle density leakage with different values of $J_y$}}
{As discussed in the main text, the particle density leakage could be reduced by choosing $J_y/J_x <1$. 
To be explicit, Fig.~\ref{fig:diff_Jy} shows the total number of particles on the edge at time $t$ for $J_y=0.8J$ and $1.0J$ in the interacting FCI phase with $J_x$ being fixed at $1.0J$.
We use the protocol in which one particle is added at the edge at $t=0$ and then time-evolve the state.
The leakage for $J_y=0.8J$ is reduced compared to the case with $J_y=1.0J$.
}
%
%---------figure------------------
%----------------------------------

\section{Further discussions on the density leakage}

%---------figure------------------
\begin{figure}
 \includegraphics[scale=0.8]{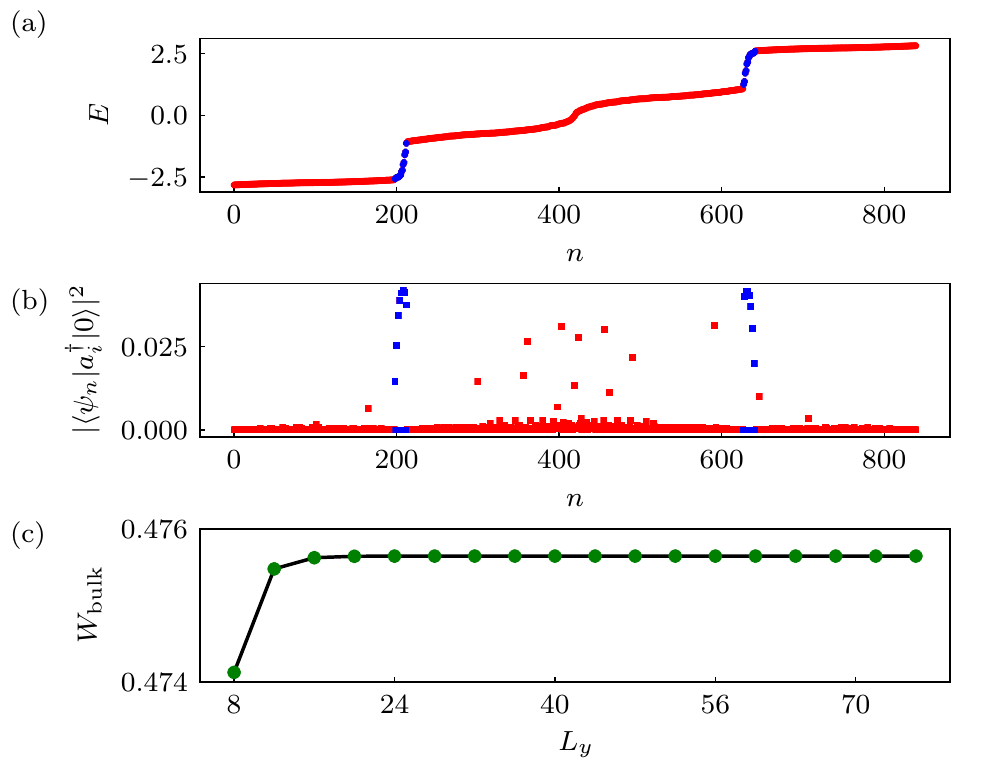}
 \caption{(a) The energy spectrum of the single-particle Hamiltonian on a cylinder with circumference $L_x=21$ and finite length $L_y=40$. The red (blue) points correspond to the bulk (edge) modes. (b) Overlap between the initial state $|\psi_0\rangle=a^{\dag}_i|0\rangle$, where $i$ is a site on the edge, and the eigenstates of the Hamiltonian $|\psi_n\rangle$. (c) The sum of all weights between $|\psi_0\rangle$ and the bulk modes for different cylinder length $L_y$.
 \label{fig:overlap}}
\end{figure}
%----------------------------------

In the non interacting case the bulk states have a finite overlap with the physical edge (i.e., the first row of sites) which are behind the density leakage. To proof this, we performed an additional numerical calculation for the non-interacting system in which we obtain the integrated overlap 
\begin{equation}
W_{\text{bulk}} = \sum_{n \in \text{bulk}} |\langle\psi_n|a^{\dag}_i|0\rangle|^2
\end{equation}
for a cylinder of length $L_y$, where $a^{\dag}_i$ creates a particle at the edge. As shown in Fig.~\ref{fig:overlap}, the weight $W_{\text{bulk}}$ increases to a finite value as the length of the cylinder gets longer. This explains the leakage into the bulk and also demonstrates that a finite but reduced density will remain at the edges at long times. To create a state that is perfectly localized, we would have to create a quasi particle at the edge which is a superposition of physical states near the edge. While we did not perform an additional simulation for the FCI, we do expect a similar finite support of bulk states on the edge.


\begin{thebibliography}{60}%
\makeatletter
\providecommand \@ifxundefined [1]{%
 \@ifx{#1\undefined}
}%
\providecommand \@ifnum [1]{%
 \ifnum #1\expandafter \@firstoftwo
 \else \expandafter \@secondoftwo
 \fi
}%
\providecommand \@ifx [1]{%
 \ifx #1\expandafter \@firstoftwo
 \else \expandafter \@secondoftwo
 \fi
}%
\providecommand \natexlab [1]{#1}%
\providecommand \enquote  [1]{``#1''}%
\providecommand \bibnamefont  [1]{#1}%
\providecommand \bibfnamefont [1]{#1}%
\providecommand \citenamefont [1]{#1}%
\providecommand \href@noop [0]{\@secondoftwo}%
\providecommand \href [0]{\begingroup \@sanitize@url \@href}%
\providecommand \@href[1]{\@@startlink{#1}\@@href}%
\providecommand \@@href[1]{\endgroup#1\@@endlink}%
\providecommand \@sanitize@url [0]{\catcode `\\12\catcode `\$12\catcode
  `\&12\catcode `\#12\catcode `\^12\catcode `\_12\catcode `\%12\relax}%
\providecommand \@@startlink[1]{}%
\providecommand \@@endlink[0]{}%
\providecommand \url  [0]{\begingroup\@sanitize@url \@url }%
\providecommand \@url [1]{\endgroup\@href {#1}{\urlprefix }}%
\providecommand \urlprefix  [0]{URL }%
\providecommand \Eprint [0]{\href }%
\providecommand \doibase [0]{http://dx.doi.org/}%
\providecommand \selectlanguage [0]{\@gobble}%
\providecommand \bibinfo  [0]{\@secondoftwo}%
\providecommand \bibfield  [0]{\@secondoftwo}%
\providecommand \translation [1]{[#1]}%
\providecommand \BibitemOpen [0]{}%
\providecommand \bibitemStop [0]{}%
\providecommand \bibitemNoStop [0]{.\EOS\space}%
\providecommand \EOS [0]{\spacefactor3000\relax}%
\providecommand \BibitemShut  [1]{\csname bibitem#1\endcsname}%
\let\auto@bib@innerbib\@empty
%</preamble>
\bibitem [{\citenamefont {Polkovnikov}\ \emph {et~al.}(2011)\citenamefont
  {Polkovnikov}, \citenamefont {Sengupta}, \citenamefont {Silva},\ and\
  \citenamefont {Vengalattore}}]{Polkovnikov2011}%
  \BibitemOpen
  \bibfield  {author} {\bibinfo {author} {\bibfnamefont {Anatoli}\ \bibnamefont
  {Polkovnikov}}, \bibinfo {author} {\bibfnamefont {Krishnendu}\ \bibnamefont
  {Sengupta}}, \bibinfo {author} {\bibfnamefont {Alessandro}\ \bibnamefont
  {Silva}}, \ and\ \bibinfo {author} {\bibfnamefont {Mukund}\ \bibnamefont
  {Vengalattore}},\ }\bibfield  {title} {\enquote {\bibinfo {title}
  {{Colloquium: Nonequilibrium dynamics of closed interacting quantum
  systems}},}\ }\href {\doibase 10.1103/RevModPhys.83.863} {\bibfield
  {journal} {\bibinfo  {journal} {Rev. Mod. Phys.}\ }\textbf {\bibinfo {volume}
  {83}},\ \bibinfo {pages} {863--883} (\bibinfo {year} {2011})}\BibitemShut
  {NoStop}%
\bibitem [{\citenamefont {Eisert}\ \emph {et~al.}(2015)\citenamefont {Eisert},
  \citenamefont {Friesdorf},\ and\ \citenamefont {Gogolin}}]{Eisert2015}%
  \BibitemOpen
  \bibfield  {author} {\bibinfo {author} {\bibfnamefont {J.}~\bibnamefont
  {Eisert}}, \bibinfo {author} {\bibfnamefont {M.}~\bibnamefont {Friesdorf}}, \
  and\ \bibinfo {author} {\bibfnamefont {C.}~\bibnamefont {Gogolin}},\
  }\bibfield  {title} {\enquote {\bibinfo {title} {Quantum many-body systems
  out of equilibrium},}\ }\href {http://dx.doi.org/10.1038/nphys3215}
  {\bibfield  {journal} {\bibinfo  {journal} {Nat Phys}\ }\textbf {\bibinfo
  {volume} {11}},\ \bibinfo {pages} {124--130} (\bibinfo {year}
  {2015})}\BibitemShut {NoStop}%
\bibitem [{\citenamefont {Bloch}(2005)}]{Bloch2005}%
  \BibitemOpen
  \bibfield  {author} {\bibinfo {author} {\bibfnamefont {Immanuel}\
  \bibnamefont {Bloch}},\ }\bibfield  {title} {\enquote {\bibinfo {title}
  {{Ultracold quantum gases in optical lattices}},}\ }\href {\doibase
  10.1038/nphys138} {\bibfield  {journal} {\bibinfo  {journal} {Nature
  Physics}\ }\textbf {\bibinfo {volume} {1}},\ \bibinfo {pages} {23--30}
  (\bibinfo {year} {2005})},\ \Eprint {http://arxiv.org/abs/0912.3646}
  {arXiv:0912.3646} \BibitemShut {NoStop}%
\bibitem [{\citenamefont {Trotzky}\ \emph {et~al.}(2012)\citenamefont
  {Trotzky}, \citenamefont {Chen}, \citenamefont {Flesch}, \citenamefont
  {McCulloch}, \citenamefont {Schollwock}, \citenamefont {Eisert},\ and\
  \citenamefont {Bloch}}]{Trotzky2012}%
  \BibitemOpen
  \bibfield  {author} {\bibinfo {author} {\bibfnamefont {S.}~\bibnamefont
  {Trotzky}}, \bibinfo {author} {\bibfnamefont {Y.-A.}\ \bibnamefont {Chen}},
  \bibinfo {author} {\bibfnamefont {A.}~\bibnamefont {Flesch}}, \bibinfo
  {author} {\bibfnamefont {I.~P.}\ \bibnamefont {McCulloch}}, \bibinfo {author}
  {\bibfnamefont {U.}~\bibnamefont {Schollwock}}, \bibinfo {author}
  {\bibfnamefont {J.}~\bibnamefont {Eisert}}, \ and\ \bibinfo {author}
  {\bibfnamefont {I.}~\bibnamefont {Bloch}},\ }\bibfield  {title} {\enquote
  {\bibinfo {title} {{Probing the relaxation towards equilibrium in an isolated
  strongly correlated one-dimensional Bose gas}},}\ }\href {\doibase
  10.1038/nphys2232} {\bibfield  {journal} {\bibinfo  {journal} {Nat Phys}\
  }\textbf {\bibinfo {volume} {8}},\ \bibinfo {pages} {325--330} (\bibinfo
  {year} {2012})}\BibitemShut {NoStop}%
\bibitem [{\citenamefont {Schreiber}\ \emph {et~al.}(2015)\citenamefont
  {Schreiber}, \citenamefont {Hodgman}, \citenamefont {Bordia}, \citenamefont
  {L{\"u}schen}, \citenamefont {Fischer}, \citenamefont {Vosk}, \citenamefont
  {Altman}, \citenamefont {Schneider},\ and\ \citenamefont
  {Bloch}}]{Schreiber2015}%
  \BibitemOpen
  \bibfield  {author} {\bibinfo {author} {\bibfnamefont {Michael}\ \bibnamefont
  {Schreiber}}, \bibinfo {author} {\bibfnamefont {Sean~S.}\ \bibnamefont
  {Hodgman}}, \bibinfo {author} {\bibfnamefont {Pranjal}\ \bibnamefont
  {Bordia}}, \bibinfo {author} {\bibfnamefont {Henrik~P.}\ \bibnamefont
  {L{\"u}schen}}, \bibinfo {author} {\bibfnamefont {Mark~H.}\ \bibnamefont
  {Fischer}}, \bibinfo {author} {\bibfnamefont {Ronen}\ \bibnamefont {Vosk}},
  \bibinfo {author} {\bibfnamefont {Ehud}\ \bibnamefont {Altman}}, \bibinfo
  {author} {\bibfnamefont {Ulrich}\ \bibnamefont {Schneider}}, \ and\ \bibinfo
  {author} {\bibfnamefont {Immanuel}\ \bibnamefont {Bloch}},\ }\bibfield
  {title} {\enquote {\bibinfo {title} {Observation of many-body localization of
  interacting fermions in a quasirandom optical lattice},}\ }\href {\doibase
  10.1126/science.aaa7432} {\bibfield  {journal} {\bibinfo  {journal}
  {Science}\ }\textbf {\bibinfo {volume} {349}},\ \bibinfo {pages} {842--845}
  (\bibinfo {year} {2015})}\BibitemShut {NoStop}%
\bibitem [{\citenamefont {Choi}\ \emph {et~al.}(2016)\citenamefont {Choi},
  \citenamefont {Hild}, \citenamefont {Zeiher}, \citenamefont {Schau{\ss}},
  \citenamefont {Rubio-Abadal}, \citenamefont {Yefsah}, \citenamefont
  {Khemani}, \citenamefont {Huse}, \citenamefont {Bloch},\ and\ \citenamefont
  {Gross}}]{Choi2016}%
  \BibitemOpen
  \bibfield  {author} {\bibinfo {author} {\bibfnamefont {Jae-yoon}\
  \bibnamefont {Choi}}, \bibinfo {author} {\bibfnamefont {Sebastian}\
  \bibnamefont {Hild}}, \bibinfo {author} {\bibfnamefont {Johannes}\
  \bibnamefont {Zeiher}}, \bibinfo {author} {\bibfnamefont {Peter}\
  \bibnamefont {Schau{\ss}}}, \bibinfo {author} {\bibfnamefont {Antonio}\
  \bibnamefont {Rubio-Abadal}}, \bibinfo {author} {\bibfnamefont {Tarik}\
  \bibnamefont {Yefsah}}, \bibinfo {author} {\bibfnamefont {Vedika}\
  \bibnamefont {Khemani}}, \bibinfo {author} {\bibfnamefont {David~A.}\
  \bibnamefont {Huse}}, \bibinfo {author} {\bibfnamefont {Immanuel}\
  \bibnamefont {Bloch}}, \ and\ \bibinfo {author} {\bibfnamefont {Christian}\
  \bibnamefont {Gross}},\ }\bibfield  {title} {\enquote {\bibinfo {title}
  {Exploring the many-body localization transition in two dimensions},}\ }\href
  {\doibase 10.1126/science.aaf8834} {\bibfield  {journal} {\bibinfo  {journal}
  {Science}\ }\textbf {\bibinfo {volume} {352}},\ \bibinfo {pages} {1547--1552}
  (\bibinfo {year} {2016})}\BibitemShut {NoStop}%
\bibitem [{\citenamefont {Tai}\ \emph {et~al.}(2017)\citenamefont {Tai},
  \citenamefont {Lukin}, \citenamefont {Rispoli}, \citenamefont {Schittko},
  \citenamefont {Menke}, \citenamefont {Borgnia}, \citenamefont {Preiss},
  \citenamefont {Grusdt}, \citenamefont {Kaufman},\ and\ \citenamefont
  {Greiner}}]{tai2017microscopy}%
  \BibitemOpen
  \bibfield  {author} {\bibinfo {author} {\bibfnamefont {M~Eric}\ \bibnamefont
  {Tai}}, \bibinfo {author} {\bibfnamefont {Alexander}\ \bibnamefont {Lukin}},
  \bibinfo {author} {\bibfnamefont {Matthew}\ \bibnamefont {Rispoli}}, \bibinfo
  {author} {\bibfnamefont {Robert}\ \bibnamefont {Schittko}}, \bibinfo {author}
  {\bibfnamefont {Tim}\ \bibnamefont {Menke}}, \bibinfo {author} {\bibfnamefont
  {Dan}\ \bibnamefont {Borgnia}}, \bibinfo {author} {\bibfnamefont {Philipp~M}\
  \bibnamefont {Preiss}}, \bibinfo {author} {\bibfnamefont {Fabian}\
  \bibnamefont {Grusdt}}, \bibinfo {author} {\bibfnamefont {Adam~M}\
  \bibnamefont {Kaufman}}, \ and\ \bibinfo {author} {\bibfnamefont {Markus}\
  \bibnamefont {Greiner}},\ }\bibfield  {title} {\enquote {\bibinfo {title}
  {{Microscopy of the interacting Harper--Hofstadter model in the two-body
  limit}},}\ }\href {\doibase 10.1038/nature22811} {\bibfield  {journal}
  {\bibinfo  {journal} {Nature}\ }\textbf {\bibinfo {volume} {546}},\ \bibinfo
  {pages} {519--523} (\bibinfo {year} {2017})}\BibitemShut {NoStop}%
\bibitem [{\citenamefont {{Harper}}(1955)}]{Harper1955}%
  \BibitemOpen
  \bibfield  {author} {\bibinfo {author} {\bibfnamefont {P.~G.}\ \bibnamefont
  {{Harper}}},\ }\bibfield  {title} {\enquote {\bibinfo {title} {{Single Band
  Motion of Conduction Electrons in a Uniform Magnetic Field}},}\ }\href
  {\doibase 10.1088/0370-1298/68/10/304} {\bibfield  {journal} {\bibinfo
  {journal} {Proceedings of the Physical Society A}\ }\textbf {\bibinfo
  {volume} {68}},\ \bibinfo {pages} {874--878} (\bibinfo {year}
  {1955})}\BibitemShut {NoStop}%
\bibitem [{\citenamefont {Hofstadter}(1976)}]{Hofstadter1976}%
  \BibitemOpen
  \bibfield  {author} {\bibinfo {author} {\bibfnamefont {Douglas~R.}\
  \bibnamefont {Hofstadter}},\ }\bibfield  {title} {\enquote {\bibinfo {title}
  {{Energy levels and wave functions of Bloch electrons in rational and
  irrational magnetic fields}},}\ }\href {\doibase 10.1103/PhysRevB.14.2239}
  {\bibfield  {journal} {\bibinfo  {journal} {Phys. Rev. B}\ }\textbf {\bibinfo
  {volume} {14}},\ \bibinfo {pages} {2239--2249} (\bibinfo {year}
  {1976})}\BibitemShut {NoStop}%
\bibitem [{\citenamefont {Aidelsburger}\ \emph {et~al.}(2013)\citenamefont
  {Aidelsburger}, \citenamefont {Atala}, \citenamefont {Lohse}, \citenamefont
  {Barreiro}, \citenamefont {Paredes},\ and\ \citenamefont
  {Bloch}}]{Aidelsburger2013}%
  \BibitemOpen
  \bibfield  {author} {\bibinfo {author} {\bibfnamefont {M.}~\bibnamefont
  {Aidelsburger}}, \bibinfo {author} {\bibfnamefont {M.}~\bibnamefont {Atala}},
  \bibinfo {author} {\bibfnamefont {M.}~\bibnamefont {Lohse}}, \bibinfo
  {author} {\bibfnamefont {J.~T.}\ \bibnamefont {Barreiro}}, \bibinfo {author}
  {\bibfnamefont {B.}~\bibnamefont {Paredes}}, \ and\ \bibinfo {author}
  {\bibfnamefont {I.}~\bibnamefont {Bloch}},\ }\bibfield  {title} {\enquote
  {\bibinfo {title} {{Realization of the Hofstadter Hamiltonian with Ultracold
  Atoms in Optical Lattices}},}\ }\href {\doibase
  10.1103/PhysRevLett.111.185301} {\bibfield  {journal} {\bibinfo  {journal}
  {Phys. Rev. Lett.}\ }\textbf {\bibinfo {volume} {111}},\ \bibinfo {pages}
  {185301} (\bibinfo {year} {2013})}\BibitemShut {NoStop}%
\bibitem [{\citenamefont {Miyake}\ \emph {et~al.}(2013)\citenamefont {Miyake},
  \citenamefont {Siviloglou}, \citenamefont {Kennedy}, \citenamefont {Burton},\
  and\ \citenamefont {Ketterle}}]{Miyake2013}%
  \BibitemOpen
  \bibfield  {author} {\bibinfo {author} {\bibfnamefont {Hirokazu}\
  \bibnamefont {Miyake}}, \bibinfo {author} {\bibfnamefont {Georgios~A.}\
  \bibnamefont {Siviloglou}}, \bibinfo {author} {\bibfnamefont {Colin~J.}\
  \bibnamefont {Kennedy}}, \bibinfo {author} {\bibfnamefont {William~Cody}\
  \bibnamefont {Burton}}, \ and\ \bibinfo {author} {\bibfnamefont {Wolfgang}\
  \bibnamefont {Ketterle}},\ }\bibfield  {title} {\enquote {\bibinfo {title}
  {{Realizing the Harper Hamiltonian with Laser-Assisted Tunneling in Optical
  Lattices}},}\ }\href {\doibase 10.1103/PhysRevLett.111.185302} {\bibfield
  {journal} {\bibinfo  {journal} {Phys. Rev. Lett.}\ }\textbf {\bibinfo
  {volume} {111}},\ \bibinfo {pages} {185302} (\bibinfo {year}
  {2013})}\BibitemShut {NoStop}%
\bibitem [{\citenamefont {Haldane}(1988)}]{Haldane1988}%
  \BibitemOpen
  \bibfield  {author} {\bibinfo {author} {\bibfnamefont {F.~D.~M.}\
  \bibnamefont {Haldane}},\ }\bibfield  {title} {\enquote {\bibinfo {title}
  {{Model for a Quantum Hall Effect without Landau Levels: Condensed-Matter
  Realization of the "Parity Anomaly"}},}\ }\href {\doibase
  10.1103/PhysRevLett.61.2015} {\bibfield  {journal} {\bibinfo  {journal}
  {Phys. Rev. Lett.}\ }\textbf {\bibinfo {volume} {61}},\ \bibinfo {pages}
  {2015--2018} (\bibinfo {year} {1988})}\BibitemShut {NoStop}%
\bibitem [{\citenamefont {Jotzu}\ \emph {et~al.}(2014)\citenamefont {Jotzu},
  \citenamefont {Messer}, \citenamefont {Desbuquois}, \citenamefont {Lebrat},
  \citenamefont {Uehlinger}, \citenamefont {Greif},\ and\ \citenamefont
  {Esslinger}}]{Jotzu2014}%
  \BibitemOpen
  \bibfield  {author} {\bibinfo {author} {\bibfnamefont {Gregor}\ \bibnamefont
  {Jotzu}}, \bibinfo {author} {\bibfnamefont {Michael}\ \bibnamefont {Messer}},
  \bibinfo {author} {\bibfnamefont {Remi}\ \bibnamefont {Desbuquois}}, \bibinfo
  {author} {\bibfnamefont {Martin}\ \bibnamefont {Lebrat}}, \bibinfo {author}
  {\bibfnamefont {Thomas}\ \bibnamefont {Uehlinger}}, \bibinfo {author}
  {\bibfnamefont {Daniel}\ \bibnamefont {Greif}}, \ and\ \bibinfo {author}
  {\bibfnamefont {Tilman}\ \bibnamefont {Esslinger}},\ }\bibfield  {title}
  {\enquote {\bibinfo {title} {{Experimental realization of the topological
  Haldane model with ultracold fermions}},}\ }\href {\doibase
  10.1038/nature13915} {\bibfield  {journal} {\bibinfo  {journal} {Nature}\
  }\textbf {\bibinfo {volume} {515}},\ \bibinfo {pages} {237--240} (\bibinfo
  {year} {2014})}\BibitemShut {NoStop}%
\bibitem [{\citenamefont {Aidelsburger}\ \emph {et~al.}(2014)\citenamefont
  {Aidelsburger}, \citenamefont {Lohse}, \citenamefont {Schweizer},
  \citenamefont {Atala}, \citenamefont {Barreiro}, \citenamefont {Nascimbene},
  \citenamefont {Cooper}, \citenamefont {Bloch},\ and\ \citenamefont
  {Goldman}}]{Aidelsburger2014}%
  \BibitemOpen
  \bibfield  {author} {\bibinfo {author} {\bibfnamefont {Monika}\ \bibnamefont
  {Aidelsburger}}, \bibinfo {author} {\bibfnamefont {Michael}\ \bibnamefont
  {Lohse}}, \bibinfo {author} {\bibfnamefont {C}~\bibnamefont {Schweizer}},
  \bibinfo {author} {\bibfnamefont {Marcos}\ \bibnamefont {Atala}}, \bibinfo
  {author} {\bibfnamefont {Julio~T}\ \bibnamefont {Barreiro}}, \bibinfo
  {author} {\bibfnamefont {S}~\bibnamefont {Nascimbene}}, \bibinfo {author}
  {\bibfnamefont {NR}~\bibnamefont {Cooper}}, \bibinfo {author} {\bibfnamefont
  {Immanuel}\ \bibnamefont {Bloch}}, \ and\ \bibinfo {author} {\bibfnamefont
  {N}~\bibnamefont {Goldman}},\ }\bibfield  {title} {\enquote {\bibinfo {title}
  {{Measuring the Chern number of Hofstadter bands with ultracold bosonic
  atoms}},}\ }\href@noop {} {\bibfield  {journal} {\bibinfo  {journal} {Nature
  Physics}\ } (\bibinfo {year} {2014})}\BibitemShut {NoStop}%
\bibitem [{\citenamefont {Jaksch}\ and\ \citenamefont
  {Zoller}(2003)}]{Jaksch2003}%
  \BibitemOpen
  \bibfield  {author} {\bibinfo {author} {\bibfnamefont {D}~\bibnamefont
  {Jaksch}}\ and\ \bibinfo {author} {\bibfnamefont {P}~\bibnamefont {Zoller}},\
  }\bibfield  {title} {\enquote {\bibinfo {title} {{Creation of effective
  magnetic fields in optical lattices: the Hofstadter butterfly for cold
  neutral atoms}},}\ }\href {http://stacks.iop.org/1367-2630/5/i=1/a=356}
  {\bibfield  {journal} {\bibinfo  {journal} {New Journal of Physics}\ }\textbf
  {\bibinfo {volume} {5}},\ \bibinfo {pages} {56} (\bibinfo {year}
  {2003})}\BibitemShut {NoStop}%
\bibitem [{\citenamefont {Rahav}\ \emph {et~al.}(2003)\citenamefont {Rahav},
  \citenamefont {Gilary},\ and\ \citenamefont {Fishman}}]{Rahav2003}%
  \BibitemOpen
  \bibfield  {author} {\bibinfo {author} {\bibfnamefont {Saar}\ \bibnamefont
  {Rahav}}, \bibinfo {author} {\bibfnamefont {Ido}\ \bibnamefont {Gilary}}, \
  and\ \bibinfo {author} {\bibfnamefont {Shmuel}\ \bibnamefont {Fishman}},\
  }\bibfield  {title} {\enquote {\bibinfo {title} {{Effective Hamiltonians for
  periodically driven systems}},}\ }\href {\doibase 10.1103/PhysRevA.68.013820}
  {\bibfield  {journal} {\bibinfo  {journal} {Phys. Rev. A}\ }\textbf {\bibinfo
  {volume} {68}},\ \bibinfo {pages} {013820} (\bibinfo {year}
  {2003})}\BibitemShut {NoStop}%
\bibitem [{\citenamefont {Dalibard}\ \emph {et~al.}(2011)\citenamefont
  {Dalibard}, \citenamefont {Gerbier}, \citenamefont
  {Juzeli\ifmmode~\bar{u}\else \={u}\fi{}nas},\ and\ \citenamefont
  {\"Ohberg}}]{Dalibard2011}%
  \BibitemOpen
  \bibfield  {author} {\bibinfo {author} {\bibfnamefont {Jean}\ \bibnamefont
  {Dalibard}}, \bibinfo {author} {\bibfnamefont {Fabrice}\ \bibnamefont
  {Gerbier}}, \bibinfo {author} {\bibfnamefont {Gediminas}\ \bibnamefont
  {Juzeli\ifmmode~\bar{u}\else \={u}\fi{}nas}}, \ and\ \bibinfo {author}
  {\bibfnamefont {Patrik}\ \bibnamefont {\"Ohberg}},\ }\bibfield  {title}
  {\enquote {\bibinfo {title} {{Colloquium: Artificial gauge potentials for
  neutral atoms}},}\ }\href {\doibase 10.1103/RevModPhys.83.1523} {\bibfield
  {journal} {\bibinfo  {journal} {Rev. Mod. Phys.}\ }\textbf {\bibinfo {volume}
  {83}},\ \bibinfo {pages} {1523--1543} (\bibinfo {year} {2011})}\BibitemShut
  {NoStop}%
\bibitem [{\citenamefont {Goldman}\ \emph
  {et~al.}(2016{\natexlab{a}})\citenamefont {Goldman}, \citenamefont {Budich},\
  and\ \citenamefont {Zoller}}]{Goldman2016}%
  \BibitemOpen
  \bibfield  {author} {\bibinfo {author} {\bibfnamefont {N}~\bibnamefont
  {Goldman}}, \bibinfo {author} {\bibfnamefont {J~C}\ \bibnamefont {Budich}}, \
  and\ \bibinfo {author} {\bibfnamefont {P}~\bibnamefont {Zoller}},\ }\bibfield
   {title} {\enquote {\bibinfo {title} {{Topological quantum matter with
  ultracold gases in optical lattices}},}\ }\href {\doibase 10.1038/nphys3803}
  {\bibfield  {journal} {\bibinfo  {journal} {Nature Physics}\ }\textbf
  {\bibinfo {volume} {12}},\ \bibinfo {pages} {639--645} (\bibinfo {year}
  {2016}{\natexlab{a}})}\BibitemShut {NoStop}%
\bibitem [{\citenamefont {Eckardt}(2017)}]{Eckardt2017}%
  \BibitemOpen
  \bibfield  {author} {\bibinfo {author} {\bibfnamefont {Andr\'e}\ \bibnamefont
  {Eckardt}},\ }\bibfield  {title} {\enquote {\bibinfo {title} {{Colloquium:
  Atomic quantum gases in periodically driven optical lattices}},}\ }\href
  {\doibase 10.1103/RevModPhys.89.011004} {\bibfield  {journal} {\bibinfo
  {journal} {Rev. Mod. Phys.}\ }\textbf {\bibinfo {volume} {89}},\ \bibinfo
  {pages} {011004} (\bibinfo {year} {2017})}\BibitemShut {NoStop}%
\bibitem [{\citenamefont {Grushin}\ \emph {et~al.}(2014)\citenamefont
  {Grushin}, \citenamefont {G{\'{o}}mez-Le{\'{o}}n},\ and\ \citenamefont
  {Neupert}}]{Grushin2014}%
  \BibitemOpen
  \bibfield  {author} {\bibinfo {author} {\bibfnamefont {Adolfo~G.}\
  \bibnamefont {Grushin}}, \bibinfo {author} {\bibfnamefont {A.}~\bibnamefont
  {G{\'{o}}mez-Le{\'{o}}n}}, \ and\ \bibinfo {author} {\bibfnamefont
  {T.}~\bibnamefont {Neupert}},\ }\bibfield  {title} {\enquote {\bibinfo
  {title} {{Floquet fractional Chern insulators}},}\ }\href {\doibase
  10.1103/PhysRevLett.112.156801} {\bibfield  {journal} {\bibinfo  {journal}
  {Physical Review Letters}\ }\textbf {\bibinfo {volume} {112}} (\bibinfo
  {year} {2014}),\ 10.1103/PhysRevLett.112.156801}\BibitemShut {NoStop}%
\bibitem [{\citenamefont {Anisimovas}\ \emph {et~al.}(2015)\citenamefont
  {Anisimovas}, \citenamefont {\ifmmode~\check{Z}\else \v{Z}\fi{}labys},
  \citenamefont {Anderson}, \citenamefont {Juzeli\ifmmode~\bar{u}\else
  \={u}\fi{}nas},\ and\ \citenamefont {Eckardt}}]{PhysRevB.91.245135}%
  \BibitemOpen
  \bibfield  {author} {\bibinfo {author} {\bibfnamefont {Egidijus}\
  \bibnamefont {Anisimovas}}, \bibinfo {author} {\bibfnamefont {Giedrius}\
  \bibnamefont {\ifmmode~\check{Z}\else \v{Z}\fi{}labys}}, \bibinfo {author}
  {\bibfnamefont {Brandon~M.}\ \bibnamefont {Anderson}}, \bibinfo {author}
  {\bibfnamefont {Gediminas}\ \bibnamefont {Juzeli\ifmmode~\bar{u}\else
  \={u}\fi{}nas}}, \ and\ \bibinfo {author} {\bibfnamefont {Andr\'e}\
  \bibnamefont {Eckardt}},\ }\bibfield  {title} {\enquote {\bibinfo {title}
  {{Role of real-space micromotion for bosonic and fermionic Floquet fractional
  Chern insulators}},}\ }\href {\doibase 10.1103/PhysRevB.91.245135} {\bibfield
   {journal} {\bibinfo  {journal} {Phys. Rev. B}\ }\textbf {\bibinfo {volume}
  {91}},\ \bibinfo {pages} {245135} (\bibinfo {year} {2015})}\BibitemShut
  {NoStop}%
\bibitem [{\citenamefont {Ra\ifmmode \check{c}\else
  \v{c}\fi{}i\ifmmode~\bar{u}\else \={u}\fi{}nas}\ \emph
  {et~al.}(2016)\citenamefont {Ra\ifmmode \check{c}\else
  \v{c}\fi{}i\ifmmode~\bar{u}\else \={u}\fi{}nas}, \citenamefont
  {\ifmmode~\check{Z}\else \v{Z}\fi{}labys}, \citenamefont {Eckardt},\ and\
  \citenamefont {Anisimovas}}]{PhysRevA.93.043618}%
  \BibitemOpen
  \bibfield  {author} {\bibinfo {author} {\bibfnamefont {Mantas}\ \bibnamefont
  {Ra\ifmmode \check{c}\else \v{c}\fi{}i\ifmmode~\bar{u}\else \={u}\fi{}nas}},
  \bibinfo {author} {\bibfnamefont {Giedrius}\ \bibnamefont
  {\ifmmode~\check{Z}\else \v{Z}\fi{}labys}}, \bibinfo {author} {\bibfnamefont
  {Andr\'e}\ \bibnamefont {Eckardt}}, \ and\ \bibinfo {author} {\bibfnamefont
  {Egidijus}\ \bibnamefont {Anisimovas}},\ }\bibfield  {title} {\enquote
  {\bibinfo {title} {{Modified interactions in a Floquet topological system on
  a square lattice and their impact on a bosonic fractional Chern insulator
  state}},}\ }\href {\doibase 10.1103/PhysRevA.93.043618} {\bibfield  {journal}
  {\bibinfo  {journal} {Phys. Rev. A}\ }\textbf {\bibinfo {volume} {93}},\
  \bibinfo {pages} {043618} (\bibinfo {year} {2016})}\BibitemShut {NoStop}%
\bibitem [{\citenamefont {Bergholtz}\ and\ \citenamefont
  {Liu}(2013)}]{BERGHOLTZ2013}%
  \BibitemOpen
  \bibfield  {author} {\bibinfo {author} {\bibfnamefont {Emil~J.}\ \bibnamefont
  {Bergholtz}}\ and\ \bibinfo {author} {\bibfnamefont {Zhao}\ \bibnamefont
  {Liu}},\ }\bibfield  {title} {\enquote {\bibinfo {title} {{Topological flat
  band models and fractional Chern insulators}},}\ }\href {\doibase
  10.1142/S021797921330017X} {\bibfield  {journal} {\bibinfo  {journal}
  {International Journal of Modern Physics B}\ }\textbf {\bibinfo {volume}
  {27}},\ \bibinfo {pages} {1330017} (\bibinfo {year} {2013})}\BibitemShut
  {NoStop}%
\bibitem [{\citenamefont {Neupert}\ \emph {et~al.}(2015)\citenamefont
  {Neupert}, \citenamefont {Chamon}, \citenamefont {Iadecola}, \citenamefont
  {Santos},\ and\ \citenamefont {Mudry}}]{Neupert2015}%
  \BibitemOpen
  \bibfield  {author} {\bibinfo {author} {\bibfnamefont {Titus}\ \bibnamefont
  {Neupert}}, \bibinfo {author} {\bibfnamefont {Claudio}\ \bibnamefont
  {Chamon}}, \bibinfo {author} {\bibfnamefont {Thomas}\ \bibnamefont
  {Iadecola}}, \bibinfo {author} {\bibfnamefont {Luiz~H}\ \bibnamefont
  {Santos}}, \ and\ \bibinfo {author} {\bibfnamefont {Christopher}\
  \bibnamefont {Mudry}},\ }\bibfield  {title} {\enquote {\bibinfo {title}
  {{Fractional (Chern and topological) insulators}},}\ }\href
  {http://iopscience.iop.org/article/10.1088/0031-8949/2015/T164/014005/pdf} {\
   (\bibinfo {year} {2015})}\BibitemShut {NoStop}%
\bibitem [{\citenamefont {Neupert}\ \emph {et~al.}(2011)\citenamefont
  {Neupert}, \citenamefont {Santos}, \citenamefont {Chamon},\ and\
  \citenamefont {Mudry}}]{Neupert2011}%
  \BibitemOpen
  \bibfield  {author} {\bibinfo {author} {\bibfnamefont {Titus}\ \bibnamefont
  {Neupert}}, \bibinfo {author} {\bibfnamefont {Luiz}\ \bibnamefont {Santos}},
  \bibinfo {author} {\bibfnamefont {Claudio}\ \bibnamefont {Chamon}}, \ and\
  \bibinfo {author} {\bibfnamefont {Christopher}\ \bibnamefont {Mudry}},\
  }\bibfield  {title} {\enquote {\bibinfo {title} {{Fractional quantum hall
  states at zero magnetic field}},}\ }\href {\doibase
  10.1103/PhysRevLett.106.236804} {\bibfield  {journal} {\bibinfo  {journal}
  {Physical Review Letters}\ }\textbf {\bibinfo {volume} {106}},\ \bibinfo
  {pages} {236804} (\bibinfo {year} {2011})},\ \Eprint
  {http://arxiv.org/abs/1012.4723} {arXiv:1012.4723} \BibitemShut {NoStop}%
\bibitem [{\citenamefont {Tang}\ \emph {et~al.}(2010)\citenamefont {Tang},
  \citenamefont {Mei},\ and\ \citenamefont {Wen}}]{Tang2010}%
  \BibitemOpen
  \bibfield  {author} {\bibinfo {author} {\bibfnamefont {Evelyn}\ \bibnamefont
  {Tang}}, \bibinfo {author} {\bibfnamefont {Jia-Wei}\ \bibnamefont {Mei}}, \
  and\ \bibinfo {author} {\bibfnamefont {Xiao-Gang}\ \bibnamefont {Wen}},\
  }\bibfield  {title} {\enquote {\bibinfo {title} {{High temperature fractional
  quantum Hall states}},}\ }\href {\doibase 10.1103/PhysRevLett.106.236802}
  {\bibfield  {journal} {\bibinfo  {journal} {Physical Review Letters}\
  }\textbf {\bibinfo {volume} {106}},\ \bibinfo {pages} {236802} (\bibinfo
  {year} {2010})},\ \Eprint {http://arxiv.org/abs/1012.2930} {arXiv:1012.2930}
  \BibitemShut {NoStop}%
\bibitem [{\citenamefont {Sun}\ \emph {et~al.}(2011)\citenamefont {Sun},
  \citenamefont {Gu}, \citenamefont {Katsura},\ and\ \citenamefont {{Das
  Sarma}}}]{Sun2011}%
  \BibitemOpen
  \bibfield  {author} {\bibinfo {author} {\bibfnamefont {Kai}\ \bibnamefont
  {Sun}}, \bibinfo {author} {\bibfnamefont {Zhengcheng}\ \bibnamefont {Gu}},
  \bibinfo {author} {\bibfnamefont {Hosho}\ \bibnamefont {Katsura}}, \ and\
  \bibinfo {author} {\bibfnamefont {S.}~\bibnamefont {{Das Sarma}}},\
  }\bibfield  {title} {\enquote {\bibinfo {title} {{Nearly flatbands with
  nontrivial topology}},}\ }\href {\doibase 10.1103/PhysRevLett.106.236803}
  {\bibfield  {journal} {\bibinfo  {journal} {Physical Review Letters}\
  }\textbf {\bibinfo {volume} {106}},\ \bibinfo {pages} {236803} (\bibinfo
  {year} {2011})},\ \Eprint {http://arxiv.org/abs/1012.5864} {arXiv:1012.5864}
  \BibitemShut {NoStop}%
\bibitem [{\citenamefont {Regnault}\ and\ \citenamefont
  {Bernevig}(2011)}]{Regnault2011}%
  \BibitemOpen
  \bibfield  {author} {\bibinfo {author} {\bibfnamefont {N.}~\bibnamefont
  {Regnault}}\ and\ \bibinfo {author} {\bibfnamefont {B.~Andrei}\ \bibnamefont
  {Bernevig}},\ }\bibfield  {title} {\enquote {\bibinfo {title} {{Fractional
  Chern Insulator}},}\ }\href {\doibase 10.1103/PhysRevX.1.021014} {\bibfield
  {journal} {\bibinfo  {journal} {Physical Review X}\ }\textbf {\bibinfo
  {volume} {1}},\ \bibinfo {pages} {021014} (\bibinfo {year}
  {2011})}\BibitemShut {NoStop}%
\bibitem [{\citenamefont {Wu}\ \emph {et~al.}(2012)\citenamefont {Wu},
  \citenamefont {Bernevig},\ and\ \citenamefont {Regnault}}]{Wu2012}%
  \BibitemOpen
  \bibfield  {author} {\bibinfo {author} {\bibfnamefont {Yang-Le}\ \bibnamefont
  {Wu}}, \bibinfo {author} {\bibfnamefont {B.~Andrei}\ \bibnamefont
  {Bernevig}}, \ and\ \bibinfo {author} {\bibfnamefont {N.}~\bibnamefont
  {Regnault}},\ }\bibfield  {title} {\enquote {\bibinfo {title} {{Zoology of
  fractional Chern insulators}},}\ }\href {\doibase 10.1103/PhysRevB.85.075116}
  {\bibfield  {journal} {\bibinfo  {journal} {Physical Review B}\ }\textbf
  {\bibinfo {volume} {85}},\ \bibinfo {pages} {075116} (\bibinfo {year}
  {2012})}\BibitemShut {NoStop}%
\bibitem [{\citenamefont {Hafezi}\ \emph {et~al.}(2007)\citenamefont {Hafezi},
  \citenamefont {S\o{}rensen}, \citenamefont {Demler},\ and\ \citenamefont
  {Lukin}}]{Hafezi2007}%
  \BibitemOpen
  \bibfield  {author} {\bibinfo {author} {\bibfnamefont {M.}~\bibnamefont
  {Hafezi}}, \bibinfo {author} {\bibfnamefont {A.~S.}\ \bibnamefont
  {S\o{}rensen}}, \bibinfo {author} {\bibfnamefont {E.}~\bibnamefont {Demler}},
  \ and\ \bibinfo {author} {\bibfnamefont {M.~D.}\ \bibnamefont {Lukin}},\
  }\bibfield  {title} {\enquote {\bibinfo {title} {{Fractional quantum Hall
  effect in optical lattices}},}\ }\href {\doibase 10.1103/PhysRevA.76.023613}
  {\bibfield  {journal} {\bibinfo  {journal} {Phys. Rev. A}\ }\textbf {\bibinfo
  {volume} {76}},\ \bibinfo {pages} {023613} (\bibinfo {year}
  {2007})}\BibitemShut {NoStop}%
\bibitem [{\citenamefont {M\"oller}\ and\ \citenamefont
  {Cooper}(2009)}]{Moeller2009}%
  \BibitemOpen
  \bibfield  {author} {\bibinfo {author} {\bibfnamefont {G.}~\bibnamefont
  {M\"oller}}\ and\ \bibinfo {author} {\bibfnamefont {N.~R.}\ \bibnamefont
  {Cooper}},\ }\bibfield  {title} {\enquote {\bibinfo {title} {{Composite
  Fermion Theory for Bosonic Quantum Hall States on Lattices}},}\ }\href
  {\doibase 10.1103/PhysRevLett.103.105303} {\bibfield  {journal} {\bibinfo
  {journal} {Phys. Rev. Lett.}\ }\textbf {\bibinfo {volume} {103}},\ \bibinfo
  {pages} {105303} (\bibinfo {year} {2009})}\BibitemShut {NoStop}%
\bibitem [{\citenamefont {He}\ \emph {et~al.}(2017)\citenamefont {He},
  \citenamefont {Grusdt}, \citenamefont {Kaufman}, \citenamefont {Greiner},\
  and\ \citenamefont {Vishwanath}}]{He2017}%
  \BibitemOpen
  \bibfield  {author} {\bibinfo {author} {\bibfnamefont {Yin-Chen}\
  \bibnamefont {He}}, \bibinfo {author} {\bibfnamefont {Fabian}\ \bibnamefont
  {Grusdt}}, \bibinfo {author} {\bibfnamefont {Adam}\ \bibnamefont {Kaufman}},
  \bibinfo {author} {\bibfnamefont {Markus}\ \bibnamefont {Greiner}}, \ and\
  \bibinfo {author} {\bibfnamefont {Ashvin}\ \bibnamefont {Vishwanath}},\
  }\bibfield  {title} {\enquote {\bibinfo {title} {{Realizing and Adiabatically
  Preparing Bosonic Integer and Fractional Quantum Hall states in Optical
  Lattices}},}\ }\href@noop {} {\bibfield  {journal} {\bibinfo  {journal}
  {arXiv:1703.00430}\ } (\bibinfo {year} {2017})}\BibitemShut {NoStop}%
\bibitem [{\citenamefont {Gerster}\ \emph {et~al.}(2017)\citenamefont
  {Gerster}, \citenamefont {Rizzi}, \citenamefont {Silvi}, \citenamefont
  {Dalmonte},\ and\ \citenamefont {Montangero}}]{Gerster2017}%
  \BibitemOpen
  \bibfield  {author} {\bibinfo {author} {\bibfnamefont {Matthias}\
  \bibnamefont {Gerster}}, \bibinfo {author} {\bibfnamefont {Matteo}\
  \bibnamefont {Rizzi}}, \bibinfo {author} {\bibfnamefont {Pietro}\
  \bibnamefont {Silvi}}, \bibinfo {author} {\bibfnamefont {Marcello}\
  \bibnamefont {Dalmonte}}, \ and\ \bibinfo {author} {\bibfnamefont {Simone}\
  \bibnamefont {Montangero}},\ }\bibfield  {title} {\enquote {\bibinfo {title}
  {{Fractional quantum Hall effect in the interacting Hofstadter model via
  tensor networks}},}\ }\href@noop {} {\bibfield  {journal} {\bibinfo
  {journal} {arXiv:1705.06515}\ } (\bibinfo {year} {2017})}\BibitemShut
  {NoStop}%
\bibitem [{\citenamefont {Motruk}\ and\ \citenamefont
  {Pollmann}(2017)}]{Motruk2017}%
  \BibitemOpen
  \bibfield  {author} {\bibinfo {author} {\bibfnamefont {Johannes}\
  \bibnamefont {Motruk}}\ and\ \bibinfo {author} {\bibfnamefont {Frank}\
  \bibnamefont {Pollmann}},\ }\bibfield  {title} {\enquote {\bibinfo {title}
  {{Phase transitions and adiabatic preparation of a fractional Chern insulator
  in a bosonic cold atom model}},}\ }\href@noop {} {\bibfield  {journal}
  {\bibinfo  {journal} {arXiv:1707.01100}\ } (\bibinfo {year}
  {2017})}\BibitemShut {NoStop}%
\bibitem [{\citenamefont {H\"ugel}\ \emph {et~al.}(2017)\citenamefont
  {H\"ugel}, \citenamefont {Strand}, \citenamefont {Werner},\ and\
  \citenamefont {Pollet}}]{Hugel2017}%
  \BibitemOpen
  \bibfield  {author} {\bibinfo {author} {\bibfnamefont {Dario}\ \bibnamefont
  {H\"ugel}}, \bibinfo {author} {\bibfnamefont {Hugo U.~R.}\ \bibnamefont
  {Strand}}, \bibinfo {author} {\bibfnamefont {Philipp}\ \bibnamefont
  {Werner}}, \ and\ \bibinfo {author} {\bibfnamefont {Lode}\ \bibnamefont
  {Pollet}},\ }\bibfield  {title} {\enquote {\bibinfo {title} {{Anisotropic
  Harper-Hofstadter-Mott model: Competition between condensation and magnetic
  fields}},}\ }\href {\doibase 10.1103/PhysRevB.96.054431} {\bibfield
  {journal} {\bibinfo  {journal} {Phys. Rev. B}\ }\textbf {\bibinfo {volume}
  {96}},\ \bibinfo {pages} {054431} (\bibinfo {year} {2017})}\BibitemShut
  {NoStop}%
\bibitem [{\citenamefont {S\o{}rensen}\ \emph {et~al.}(2005)\citenamefont
  {S\o{}rensen}, \citenamefont {Demler},\ and\ \citenamefont
  {Lukin}}]{Sorensen2005}%
  \BibitemOpen
  \bibfield  {author} {\bibinfo {author} {\bibfnamefont {Anders~S.}\
  \bibnamefont {S\o{}rensen}}, \bibinfo {author} {\bibfnamefont {Eugene}\
  \bibnamefont {Demler}}, \ and\ \bibinfo {author} {\bibfnamefont {Mikhail~D.}\
  \bibnamefont {Lukin}},\ }\bibfield  {title} {\enquote {\bibinfo {title}
  {{Fractional Quantum Hall States of Atoms in Optical Lattices}},}\ }\href
  {\doibase 10.1103/PhysRevLett.94.086803} {\bibfield  {journal} {\bibinfo
  {journal} {Phys. Rev. Lett.}\ }\textbf {\bibinfo {volume} {94}},\ \bibinfo
  {pages} {086803} (\bibinfo {year} {2005})}\BibitemShut {NoStop}%
\bibitem [{\citenamefont {Spielman}(2013)}]{Spielman2013}%
  \BibitemOpen
  \bibfield  {author} {\bibinfo {author} {\bibfnamefont {Ian~B}\ \bibnamefont
  {Spielman}},\ }\bibfield  {title} {\enquote {\bibinfo {title} {Detection of
  topological matter with quantum gases},}\ }\href@noop {} {\bibfield
  {journal} {\bibinfo  {journal} {Annalen der Physik}\ }\textbf {\bibinfo
  {volume} {525}},\ \bibinfo {pages} {797--807} (\bibinfo {year}
  {2013})}\BibitemShut {NoStop}%
\bibitem [{\citenamefont {Goldman}\ \emph
  {et~al.}(2013{\natexlab{a}})\citenamefont {Goldman}, \citenamefont
  {Dalibard}, \citenamefont {Dauphin}, \citenamefont {Gerbier}, \citenamefont
  {Lewenstein}, \citenamefont {Zoller},\ and\ \citenamefont
  {Spielman}}]{Goldman2013}%
  \BibitemOpen
  \bibfield  {author} {\bibinfo {author} {\bibfnamefont {Nathan}\ \bibnamefont
  {Goldman}}, \bibinfo {author} {\bibfnamefont {Jean}\ \bibnamefont
  {Dalibard}}, \bibinfo {author} {\bibfnamefont {Alexandre}\ \bibnamefont
  {Dauphin}}, \bibinfo {author} {\bibfnamefont {Fabrice}\ \bibnamefont
  {Gerbier}}, \bibinfo {author} {\bibfnamefont {Maciej}\ \bibnamefont
  {Lewenstein}}, \bibinfo {author} {\bibfnamefont {Peter}\ \bibnamefont
  {Zoller}}, \ and\ \bibinfo {author} {\bibfnamefont {Ian~B}\ \bibnamefont
  {Spielman}},\ }\bibfield  {title} {\enquote {\bibinfo {title} {Direct imaging
  of topological edge states in cold-atom systems},}\ }\href@noop {} {\bibfield
   {journal} {\bibinfo  {journal} {Proceedings of the National Academy of
  Sciences}\ }\textbf {\bibinfo {volume} {110}},\ \bibinfo {pages} {6736--6741}
  (\bibinfo {year} {2013}{\natexlab{a}})}\BibitemShut {NoStop}%
\bibitem [{\citenamefont {Goldman}\ \emph {et~al.}(2012)\citenamefont
  {Goldman}, \citenamefont {Beugnon},\ and\ \citenamefont
  {Gerbier}}]{Goldman2012}%
  \BibitemOpen
  \bibfield  {author} {\bibinfo {author} {\bibfnamefont {Nathan}\ \bibnamefont
  {Goldman}}, \bibinfo {author} {\bibfnamefont {J\'er\^ome}\ \bibnamefont
  {Beugnon}}, \ and\ \bibinfo {author} {\bibfnamefont {Fabrice}\ \bibnamefont
  {Gerbier}},\ }\bibfield  {title} {\enquote {\bibinfo {title} {{Detecting
  Chiral Edge States in the Hofstadter Optical Lattice}},}\ }\href {\doibase
  10.1103/PhysRevLett.108.255303} {\bibfield  {journal} {\bibinfo  {journal}
  {Phys. Rev. Lett.}\ }\textbf {\bibinfo {volume} {108}},\ \bibinfo {pages}
  {255303} (\bibinfo {year} {2012})}\BibitemShut {NoStop}%
\bibitem [{\citenamefont {Zaletel}\ \emph {et~al.}(2015)\citenamefont
  {Zaletel}, \citenamefont {Mong}, \citenamefont {Karrasch}, \citenamefont
  {Moore},\ and\ \citenamefont {Pollmann}}]{Zaletel2015}%
  \BibitemOpen
  \bibfield  {author} {\bibinfo {author} {\bibfnamefont {Michael~P.}\
  \bibnamefont {Zaletel}}, \bibinfo {author} {\bibfnamefont {Roger S.~K.}\
  \bibnamefont {Mong}}, \bibinfo {author} {\bibfnamefont {Christoph}\
  \bibnamefont {Karrasch}}, \bibinfo {author} {\bibfnamefont {Joel~E.}\
  \bibnamefont {Moore}}, \ and\ \bibinfo {author} {\bibfnamefont {Frank}\
  \bibnamefont {Pollmann}},\ }\bibfield  {title} {\enquote {\bibinfo {title}
  {{Time-evolving a matrix product state with long-ranged interactions}},}\
  }\href {\doibase 10.1103/PhysRevB.91.165112} {\bibfield  {journal} {\bibinfo
  {journal} {Phys. Rev. B}\ }\textbf {\bibinfo {volume} {91}},\ \bibinfo
  {pages} {165112} (\bibinfo {year} {2015})}\BibitemShut {NoStop}%
\bibitem [{\citenamefont {Gohlke}\ \emph {et~al.}(2017)\citenamefont {Gohlke},
  \citenamefont {Verresen}, \citenamefont {Moessner},\ and\ \citenamefont
  {Pollmann}}]{Gohlke2017}%
  \BibitemOpen
  \bibfield  {author} {\bibinfo {author} {\bibfnamefont {Matthias}\
  \bibnamefont {Gohlke}}, \bibinfo {author} {\bibfnamefont {Ruben}\
  \bibnamefont {Verresen}}, \bibinfo {author} {\bibfnamefont {Roderich}\
  \bibnamefont {Moessner}}, \ and\ \bibinfo {author} {\bibfnamefont {Frank}\
  \bibnamefont {Pollmann}},\ }\bibfield  {title} {\enquote {\bibinfo {title}
  {{Dynamics of the Kitaev-Heisenberg Model}},}\ }\href@noop {} {\bibfield
  {journal} {\bibinfo  {journal} {arXiv:1701.04678}\ } (\bibinfo {year}
  {2017})}\BibitemShut {NoStop}%
\bibitem [{SM()}]{SM}%
  \BibitemOpen
  \href@noop {} {}\bibinfo {note} {See Supplementary material at URL for
  details on the model, the computation of the static properties, a summary of
  chiral Luttinger liquid theory and the real space time evolution of the edge
  state density distribution.}\BibitemShut {Stop}%
\bibitem [{\citenamefont {Cincio}\ and\ \citenamefont {Vidal}(2013)}]{CV13}%
  \BibitemOpen
  \bibfield  {author} {\bibinfo {author} {\bibfnamefont {L.}~\bibnamefont
  {Cincio}}\ and\ \bibinfo {author} {\bibfnamefont {G.}~\bibnamefont {Vidal}},\
  }\bibfield  {title} {\enquote {\bibinfo {title} {Characterizing topological
  order by studying the ground states on an infinite cylinder},}\ }\href
  {\doibase 10.1103/PhysRevLett.110.067208} {\bibfield  {journal} {\bibinfo
  {journal} {Phys. Rev. Lett.}\ }\textbf {\bibinfo {volume} {110}},\ \bibinfo
  {pages} {067208} (\bibinfo {year} {2013})}\BibitemShut {NoStop}%
\bibitem [{\citenamefont {Tagliacozzo}\ \emph {et~al.}(2008)\citenamefont
  {Tagliacozzo}, \citenamefont {de~Oliveira}, \citenamefont {Iblisdir},\ and\
  \citenamefont {Latorre}}]{Tagliacozzo2008}%
  \BibitemOpen
  \bibfield  {author} {\bibinfo {author} {\bibfnamefont {L.}~\bibnamefont
  {Tagliacozzo}}, \bibinfo {author} {\bibfnamefont {Thiago.~R.}\ \bibnamefont
  {de~Oliveira}}, \bibinfo {author} {\bibfnamefont {S.}~\bibnamefont
  {Iblisdir}}, \ and\ \bibinfo {author} {\bibfnamefont {J.~I.}\ \bibnamefont
  {Latorre}},\ }\bibfield  {title} {\enquote {\bibinfo {title} {Scaling of
  entanglement support for matrix product states},}\ }\href {\doibase
  10.1103/PhysRevB.78.024410} {\bibfield  {journal} {\bibinfo  {journal} {Phys.
  Rev. B}\ }\textbf {\bibinfo {volume} {78}},\ \bibinfo {pages} {024410}
  (\bibinfo {year} {2008})}\BibitemShut {NoStop}%
\bibitem [{\citenamefont {Pollmann}\ \emph {et~al.}(2009)\citenamefont
  {Pollmann}, \citenamefont {Mukerjee}, \citenamefont {Turner},\ and\
  \citenamefont {Moore}}]{Pollmann2009}%
  \BibitemOpen
  \bibfield  {author} {\bibinfo {author} {\bibfnamefont {Frank}\ \bibnamefont
  {Pollmann}}, \bibinfo {author} {\bibfnamefont {Subroto}\ \bibnamefont
  {Mukerjee}}, \bibinfo {author} {\bibfnamefont {Ari~M.}\ \bibnamefont
  {Turner}}, \ and\ \bibinfo {author} {\bibfnamefont {Joel~E.}\ \bibnamefont
  {Moore}},\ }\bibfield  {title} {\enquote {\bibinfo {title} {{Theory of
  Finite-Entanglement Scaling at One-Dimensional Quantum Critical Points}},}\
  }\href {\doibase 10.1103/PhysRevLett.102.255701} {\bibfield  {journal}
  {\bibinfo  {journal} {Phys. Rev. Lett.}\ }\textbf {\bibinfo {volume} {102}},\
  \bibinfo {pages} {255701} (\bibinfo {year} {2009})}\BibitemShut {NoStop}%
\bibitem [{\citenamefont {Calabrese}\ and\ \citenamefont
  {Cardy}(2004)}]{calabrese2004entanglement}%
  \BibitemOpen
  \bibfield  {author} {\bibinfo {author} {\bibfnamefont {Pasquale}\
  \bibnamefont {Calabrese}}\ and\ \bibinfo {author} {\bibfnamefont {John}\
  \bibnamefont {Cardy}},\ }\bibfield  {title} {\enquote {\bibinfo {title}
  {Entanglement entropy and quantum field theory},}\ }\href@noop {} {\bibfield
  {journal} {\bibinfo  {journal} {Journal of Statistical Mechanics: Theory and
  Experiment}\ }\textbf {\bibinfo {volume} {2004}},\ \bibinfo {pages} {P06002}
  (\bibinfo {year} {2004})}\BibitemShut {NoStop}%
\bibitem [{\citenamefont {Li}\ and\ \citenamefont {Haldane}(2008)}]{Li2008}%
  \BibitemOpen
  \bibfield  {author} {\bibinfo {author} {\bibfnamefont {Hui}\ \bibnamefont
  {Li}}\ and\ \bibinfo {author} {\bibfnamefont {F.~D.~M.}\ \bibnamefont
  {Haldane}},\ }\bibfield  {title} {\enquote {\bibinfo {title} {{Entanglement
  Spectrum as a Generalization of Entanglement Entropy: Identification of
  Topological Order in Non-Abelian Fractional Quantum Hall Effect States}},}\
  }\href {\doibase 10.1103/PhysRevLett.101.010504} {\bibfield  {journal}
  {\bibinfo  {journal} {Phys. Rev. Lett.}\ }\textbf {\bibinfo {volume} {101}},\
  \bibinfo {pages} {010504} (\bibinfo {year} {2008})}\BibitemShut {NoStop}%
\bibitem [{\citenamefont {Kj\"all}\ \emph {et~al.}(2011)\citenamefont
  {Kj\"all}, \citenamefont {Pollmann},\ and\ \citenamefont {Moore}}]{KPM11}%
  \BibitemOpen
  \bibfield  {author} {\bibinfo {author} {\bibfnamefont {Jonas~A.}\
  \bibnamefont {Kj\"all}}, \bibinfo {author} {\bibfnamefont {Frank}\
  \bibnamefont {Pollmann}}, \ and\ \bibinfo {author} {\bibfnamefont {Joel~E.}\
  \bibnamefont {Moore}},\ }\bibfield  {title} {\enquote {\bibinfo {title}
  {Bound states and ${E}_{8}$ symmetry effects in perturbed quantum ising
  chains},}\ }\href {\doibase 10.1103/PhysRevB.83.020407} {\bibfield  {journal}
  {\bibinfo  {journal} {Phys. Rev. B}\ }\textbf {\bibinfo {volume} {83}},\
  \bibinfo {pages} {020407} (\bibinfo {year} {2011})}\BibitemShut {NoStop}%
\bibitem [{\citenamefont {Bakr}\ \emph {et~al.}(2009)\citenamefont {Bakr},
  \citenamefont {Gillen}, \citenamefont {Peng}, \citenamefont {F{\"{o}}lling},\
  and\ \citenamefont {Greiner}}]{Bakr2009}%
  \BibitemOpen
  \bibfield  {author} {\bibinfo {author} {\bibfnamefont {Waseem~S.}\
  \bibnamefont {Bakr}}, \bibinfo {author} {\bibfnamefont {Jonathon~I.}\
  \bibnamefont {Gillen}}, \bibinfo {author} {\bibfnamefont {Amy}\ \bibnamefont
  {Peng}}, \bibinfo {author} {\bibfnamefont {Simon}\ \bibnamefont
  {F{\"{o}}lling}}, \ and\ \bibinfo {author} {\bibfnamefont {Markus}\
  \bibnamefont {Greiner}},\ }\bibfield  {title} {\enquote {\bibinfo {title} {{A
  quantum gas microscope for detecting single atoms in a Hubbard-regime optical
  lattice}},}\ }\href {\doibase 10.1038/nature08482} {\bibfield  {journal}
  {\bibinfo  {journal} {Nature}\ }\textbf {\bibinfo {volume} {462}},\ \bibinfo
  {pages} {74--77} (\bibinfo {year} {2009})}\BibitemShut {NoStop}%
\bibitem [{\citenamefont {Hilker}\ \emph {et~al.}(2017)\citenamefont {Hilker},
  \citenamefont {Salomon}, \citenamefont {Grusdt}, \citenamefont {Omran},
  \citenamefont {Boll}, \citenamefont {Demler}, \citenamefont {Bloch},\ and\
  \citenamefont {Gross}}]{Hilker484}%
  \BibitemOpen
  \bibfield  {author} {\bibinfo {author} {\bibfnamefont {Timon~A.}\
  \bibnamefont {Hilker}}, \bibinfo {author} {\bibfnamefont {Guillaume}\
  \bibnamefont {Salomon}}, \bibinfo {author} {\bibfnamefont {Fabian}\
  \bibnamefont {Grusdt}}, \bibinfo {author} {\bibfnamefont {Ahmed}\
  \bibnamefont {Omran}}, \bibinfo {author} {\bibfnamefont {Martin}\
  \bibnamefont {Boll}}, \bibinfo {author} {\bibfnamefont {Eugene}\ \bibnamefont
  {Demler}}, \bibinfo {author} {\bibfnamefont {Immanuel}\ \bibnamefont
  {Bloch}}, \ and\ \bibinfo {author} {\bibfnamefont {Christian}\ \bibnamefont
  {Gross}},\ }\bibfield  {title} {\enquote {\bibinfo {title} {{Revealing hidden
  antiferromagnetic correlations in doped Hubbard chains via string
  correlators}},}\ }\href {\doibase 10.1126/science.aam8990} {\bibfield
  {journal} {\bibinfo  {journal} {Science}\ }\textbf {\bibinfo {volume}
  {357}},\ \bibinfo {pages} {484--487} (\bibinfo {year} {2017})}\BibitemShut
  {NoStop}%
\bibitem [{\citenamefont {Wen}(1990)}]{Wen1990}%
  \BibitemOpen
  \bibfield  {author} {\bibinfo {author} {\bibfnamefont {X.~G.}\ \bibnamefont
  {Wen}},\ }\bibfield  {title} {\enquote {\bibinfo {title} {{Chiral Luttinger
  liquid and the edge excitations in the fractional quantum Hall states}},}\
  }\href {\doibase 10.1103/PhysRevB.41.12838} {\bibfield  {journal} {\bibinfo
  {journal} {Phys. Rev. B}\ }\textbf {\bibinfo {volume} {41}},\ \bibinfo
  {pages} {12838--12844} (\bibinfo {year} {1990})}\BibitemShut {NoStop}%
\bibitem [{\citenamefont {Grushin}\ \emph {et~al.}(2016)\citenamefont
  {Grushin}, \citenamefont {Roy},\ and\ \citenamefont {Haque}}]{Grushin2016}%
  \BibitemOpen
  \bibfield  {author} {\bibinfo {author} {\bibfnamefont {Adolfo~G}\
  \bibnamefont {Grushin}}, \bibinfo {author} {\bibfnamefont {Sthitadhi}\
  \bibnamefont {Roy}}, \ and\ \bibinfo {author} {\bibfnamefont {Masudul}\
  \bibnamefont {Haque}},\ }\bibfield  {title} {\enquote {\bibinfo {title}
  {{Response of fermions in Chern bands to spatially local quenches}},}\ }\href
  {http://stacks.iop.org/1742-5468/2016/i=8/a=083103} {\bibfield  {journal}
  {\bibinfo  {journal} {Journal of Statistical Mechanics: Theory and
  Experiment}\ }\textbf {\bibinfo {volume} {2016}},\ \bibinfo {pages} {083103}
  (\bibinfo {year} {2016})}\BibitemShut {NoStop}%
\bibitem [{\citenamefont {Buchhold}\ \emph {et~al.}(2012)\citenamefont
  {Buchhold}, \citenamefont {Cocks},\ and\ \citenamefont
  {Hofstetter}}]{Buchhold2012}%
  \BibitemOpen
  \bibfield  {author} {\bibinfo {author} {\bibfnamefont {Michael}\ \bibnamefont
  {Buchhold}}, \bibinfo {author} {\bibfnamefont {Daniel}\ \bibnamefont
  {Cocks}}, \ and\ \bibinfo {author} {\bibfnamefont {Walter}\ \bibnamefont
  {Hofstetter}},\ }\bibfield  {title} {\enquote {\bibinfo {title} {Effects of
  smooth boundaries on topological edge modes in optical lattices},}\ }\href
  {\doibase 10.1103/PhysRevA.85.063614} {\bibfield  {journal} {\bibinfo
  {journal} {Phys. Rev. A}\ }\textbf {\bibinfo {volume} {85}},\ \bibinfo
  {pages} {063614} (\bibinfo {year} {2012})}\BibitemShut {NoStop}%
\bibitem [{\citenamefont {Goldman}\ \emph
  {et~al.}(2013{\natexlab{b}})\citenamefont {Goldman}, \citenamefont
  {Beugnon},\ and\ \citenamefont {Gerbier}}]{Goldman2013identifying}%
  \BibitemOpen
  \bibfield  {author} {\bibinfo {author} {\bibfnamefont {Nathan}\ \bibnamefont
  {Goldman}}, \bibinfo {author} {\bibfnamefont {J{\'e}r{\^o}me}\ \bibnamefont
  {Beugnon}}, \ and\ \bibinfo {author} {\bibfnamefont {Fabrice}\ \bibnamefont
  {Gerbier}},\ }\bibfield  {title} {\enquote {\bibinfo {title} {Identifying
  topological edge states in {2D} optical lattices using light scattering},}\
  }\href@noop {} {\bibfield  {journal} {\bibinfo  {journal} {The European
  Physical Journal Special Topics}\ }\textbf {\bibinfo {volume} {217}},\
  \bibinfo {pages} {135--152} (\bibinfo {year}
  {2013}{\natexlab{b}})}\BibitemShut {NoStop}%
\bibitem [{\citenamefont {Goldman}\ \emph
  {et~al.}(2016{\natexlab{b}})\citenamefont {Goldman}, \citenamefont {Jotzu},
  \citenamefont {Messer}, \citenamefont {G\"org}, \citenamefont {Desbuquois},\
  and\ \citenamefont {Esslinger}}]{Goldman2016a}%
  \BibitemOpen
  \bibfield  {author} {\bibinfo {author} {\bibfnamefont {N.}~\bibnamefont
  {Goldman}}, \bibinfo {author} {\bibfnamefont {G.}~\bibnamefont {Jotzu}},
  \bibinfo {author} {\bibfnamefont {M.}~\bibnamefont {Messer}}, \bibinfo
  {author} {\bibfnamefont {F.}~\bibnamefont {G\"org}}, \bibinfo {author}
  {\bibfnamefont {R.}~\bibnamefont {Desbuquois}}, \ and\ \bibinfo {author}
  {\bibfnamefont {T.}~\bibnamefont {Esslinger}},\ }\bibfield  {title} {\enquote
  {\bibinfo {title} {Creating topological interfaces and detecting chiral edge
  modes in a two-dimensional optical lattice},}\ }\href {\doibase
  10.1103/PhysRevA.94.043611} {\bibfield  {journal} {\bibinfo  {journal} {Phys.
  Rev. A}\ }\textbf {\bibinfo {volume} {94}},\ \bibinfo {pages} {043611}
  (\bibinfo {year} {2016}{\natexlab{b}})}\BibitemShut {NoStop}%
\bibitem [{\citenamefont {McCulloch}(2008)}]{McCulloch2008}%
  \BibitemOpen
  \bibfield  {author} {\bibinfo {author} {\bibfnamefont {I.~P.}\ \bibnamefont
  {McCulloch}},\ }\bibfield  {title} {\enquote {\bibinfo {title} {Infinite size
  density matrix renormalization group, revisited},}\ }\href@noop {} {\bibfield
   {journal} {\bibinfo  {journal} {arXiv:0804.2509}\ } (\bibinfo {year}
  {2008})}\BibitemShut {NoStop}%
\bibitem [{\citenamefont {Kjall}\ \emph {et~al.}(2013)\citenamefont {Kjall},
  \citenamefont {Zaletel}, \citenamefont {Mong}, \citenamefont {Bardarson},\
  and\ \citenamefont {Pollmann}}]{Kjall2013}%
  \BibitemOpen
  \bibfield  {author} {\bibinfo {author} {\bibfnamefont {Jonas~A.}\
  \bibnamefont {Kjall}}, \bibinfo {author} {\bibfnamefont {Michael~P.}\
  \bibnamefont {Zaletel}}, \bibinfo {author} {\bibfnamefont {Roger S.~K.}\
  \bibnamefont {Mong}}, \bibinfo {author} {\bibfnamefont {Jens~H.}\
  \bibnamefont {Bardarson}}, \ and\ \bibinfo {author} {\bibfnamefont {Frank}\
  \bibnamefont {Pollmann}},\ }\bibfield  {title} {\enquote {\bibinfo {title}
  {{Phase diagram of the anisotropic spin-2 XXZ model: Infinite-system density
  matrix renormalization group study}},}\ }\href {\doibase
  10.1103/PhysRevB.87.235106} {\bibfield  {journal} {\bibinfo  {journal} {Phys.
  Rev. B}\ }\textbf {\bibinfo {volume} {87}},\ \bibinfo {pages} {235106}
  (\bibinfo {year} {2013})}\BibitemShut {NoStop}%
\bibitem [{\citenamefont {Laughlin}(1981)}]{Laughlin1981}%
  \BibitemOpen
  \bibfield  {author} {\bibinfo {author} {\bibfnamefont {R.~B.}\ \bibnamefont
  {Laughlin}},\ }\bibfield  {title} {\enquote {\bibinfo {title} {{Quantized
  Hall conductivity in two dimensions}},}\ }\href {\doibase
  10.1103/PhysRevB.23.5632} {\bibfield  {journal} {\bibinfo  {journal} {Phys.
  Rev. B}\ }\textbf {\bibinfo {volume} {23}},\ \bibinfo {pages} {5632--5633}
  (\bibinfo {year} {1981})}\BibitemShut {NoStop}%
\bibitem [{\citenamefont {Zaletel}\ \emph {et~al.}(2014)\citenamefont
  {Zaletel}, \citenamefont {Mong},\ and\ \citenamefont
  {Pollmann}}]{Zaletel2014}%
  \BibitemOpen
  \bibfield  {author} {\bibinfo {author} {\bibfnamefont {Michael~P}\
  \bibnamefont {Zaletel}}, \bibinfo {author} {\bibfnamefont {Roger~SK}\
  \bibnamefont {Mong}}, \ and\ \bibinfo {author} {\bibfnamefont {Frank}\
  \bibnamefont {Pollmann}},\ }\bibfield  {title} {\enquote {\bibinfo {title}
  {Flux insertion, entanglement, and quantized responses},}\ }\href@noop {}
  {\bibfield  {journal} {\bibinfo  {journal} {Journal of Statistical Mechanics:
  Theory and Experiment}\ }\textbf {\bibinfo {volume} {2014}},\ \bibinfo
  {pages} {P10007} (\bibinfo {year} {2014})}\BibitemShut {NoStop}%
\bibitem [{\citenamefont {Grushin}\ \emph {et~al.}(2015)\citenamefont
  {Grushin}, \citenamefont {Motruk}, \citenamefont {Zaletel},\ and\
  \citenamefont {Pollmann}}]{Grushin2015}%
  \BibitemOpen
  \bibfield  {author} {\bibinfo {author} {\bibfnamefont {Adolfo~G.}\
  \bibnamefont {Grushin}}, \bibinfo {author} {\bibfnamefont {Johannes}\
  \bibnamefont {Motruk}}, \bibinfo {author} {\bibfnamefont {Michael~P.}\
  \bibnamefont {Zaletel}}, \ and\ \bibinfo {author} {\bibfnamefont {Frank}\
  \bibnamefont {Pollmann}},\ }\bibfield  {title} {\enquote {\bibinfo {title}
  {{Characterization and stability of a fermionic $\ensuremath{\nu}=1/3$
  fractional Chern insulator}},}\ }\href {\doibase 10.1103/PhysRevB.91.035136}
  {\bibfield  {journal} {\bibinfo  {journal} {Phys. Rev. B}\ }\textbf {\bibinfo
  {volume} {91}},\ \bibinfo {pages} {035136} (\bibinfo {year}
  {2015})}\BibitemShut {NoStop}%
\end{thebibliography}
\end{document}